\documentclass[]{emulateapj}
\usepackage{amsmath}
\usepackage{graphicx}
\usepackage{natbib}
\usepackage{amssymb}

\usepackage{setspace}

\usepackage{color,xcolor}
\usepackage{hyperref}             
\hypersetup{pdfauthor=Christoph Mordasini}
\hypersetup{breaklinks=true,colorlinks=true,urlcolor=blue, linkcolor=blue,  citecolor=blue, bookmarksopen=true}

\definecolor{refco}{HTML}{547687}

\usepackage{etoolbox}

\makeatletter

\patchcmd{\NAT@citex}
  {\@citea\NAT@hyper@{%
     \NAT@nmfmt{\NAT@nm}%
     \hyper@natlinkbreak{\NAT@aysep\NAT@spacechar}{\@citeb\@extra@b@citeb}%
     \NAT@date}}
  {\@citea\NAT@nmfmt{\NAT@nm}%
   \NAT@aysep\NAT@spacechar\NAT@hyper@{\NAT@date}}{}{}

\patchcmd{\NAT@citex}
  {\@citea\NAT@hyper@{%
     \NAT@nmfmt{\NAT@nm}%
     \hyper@natlinkbreak{\NAT@spacechar\NAT@@open\if*#1*\else#1\NAT@spacechar\fi}%
       {\@citeb\@extra@b@citeb}%
     \NAT@date}}
  {\@citea\NAT@nmfmt{\NAT@nm}%
   \NAT@spacechar\NAT@@open\if*#1*\else#1\NAT@spacechar\fi\NAT@hyper@{\NAT@date}}
  {}{}

\makeatother

\def\tint{T_{\rm int}}

\def\teq{T_{\rm eq}}
\def\mearth{M_{\oplus}}
\def\rearth{R_{\oplus}}
\def\msun{M_{\odot}}

\begin{document}

\title{Compositional imprints in density-distance-time:  \\ a rocky composition for close-in low-mass exoplanets from the location of the valley of evaporation}

\author{Sheng Jin\altaffilmark{1} \& Christoph Mordasini\altaffilmark{2}}
\affil{$^1$
CAS Key Laboratory of Planetary Sciences, Purple Mountain Observatory,
 Chinese Academy of Sciences, Nanjing 210008, China
}

\affil{$^2$
Physikalisches Institut, Universit\"at Bern, Gesellschaftstrasse 6, 3012 Bern, Switzerland 
}

\begin{abstract}
We use an end-to-end model of planet formation, thermodynamic evolution, and atmospheric escape to investigate how the statistical imprints of evaporation depend on the bulk composition of planetary cores (rocky vs. icy). 
We find that the population-wide imprints like the location of the ``evaporation valley" in the distance--radius plane and the corresponding bimodal radius distribution clearly differ depending on the bulk composition of the cores. Comparison with the observed position of the valley (Fulton et al. 2017) suggests that close-in low-mass Kepler planets have a predominately Earth-like rocky composition. Combined with the excess of period ratios outside of MMR, this suggests that low-mass Kepler planets formed inside of the water iceline, but still undergoing orbital migration.
The core radius becomes visible for planets losing all primordial H/He. For planets in this ``triangle of evaporation" in the distance--radius plane, the degeneracy in compositions is reduced. In the observed diagram, we identify a trend to more volatile-rich compositions with increasing radius ($R/\rearth\lesssim$ 1.6 rocky; 1.6-3.0 ices and/or H/He; $\gtrsim 3$: H/He).   
The mass--density diagram contains important information about formation and evolution{. Its characteristic broken V-shape reveals} the transitions from solid planets {to} low-mass core-dominated planets with H/He {and finally} to  gas-dominated giants.
Evaporation causes density and orbital distance to be anti-correlated for low-mass planets, in contrast to giants, where closer-in planets are less dense, likely due to inflation. 
The temporal evolution of the statistical properties reported here will be of interest for the PLATO 2.0 mission which will observe the temporal dimension.
 
\end{abstract}


\section{Introduction}
The observational data on extrasolar planets has increased dramatically in the last two decades. The latest surveys conducted with different detection methods for example show that the presence of planets  is, at least around solar-like stars, the norm \citep{Mayor2011,Borucki2011,Cassan2012}. Thanks to the progress of the radial velocity and transit techniques, we have furthermore started to detect Earth-size planets in the recent years, including several planets that are potentially in the habitable zone \citep[see][]{Kopparapu2013}.

However, regarding the (geo)physical characterization of exoplanets, the information that we can observationally infer for most exoplanets is still limited to orbital elements and a minimal mass, or a radius. From a point of view of planet formation theory, a better knowledge of the basic (geo)physical properties of an exoplanet like its bulk composition is highly desirable as it is closely related to its formation history. For example, the presence of close-in low-mass planets consisting mainly of ices would indicate that these planets have formed outside of the iceline, and then migrated towards the host star. The frequency of such planets would thus serve as an important observational constraint for Type I planet migration models \citep[e.g.,][]{paardekooperbaruteau2010,dittkristmordasini2014}. This is of high interest for the currently debated formation mechanism of this frequent type of planet (in situ versus orbital migration, e.g., \citealt{chianglaughlin2013,ogiharamorbidelli2015}).  Besides this, it is also a critical factor in determining the habitability of a planet \citep[e.g.][]{Alibert2014,kitzmann2015}.

For a handful of transiting planets that have transmission or thermal emission spectra, and several direct imaging planets, one can derive constraints on their atmospheric structures and chemical composition from multi-band photometry or spectroscopy \citep[e.g.,][]{Richardson2007,Madhusudhan2011GJ436b,Konopacky2013}. But for the majority of exoplanets, it is currently not feasible to obtain the spectrum.

For a significantly higher number of exoplanets, the only (geo)physical constraints we can get, beside the orbital properties, have to be derived from the  planetary mass (from radial velocity observations) combined with the planetary radius (from transit observations). Such combined measurements directly lead to the mean density of the planet which can be used as a first constraint on the bulk composition \citep[e.g.,][]{Valencia2007,Valencia2010,Rogers2010GJ1214b,Rogers2010density}.

However, this is a relatively limited approach due to the degeneracy in the planetary mass--radius relationship \citep{Seager2007,Valencia2010}, especially for low-mass planets with a gaseous H/He envelope \citep{Rogers2010GJ1214b,Valencia2013,Howe2014}: a silicate-iron core combined with a (potentially tenuous) H/He envelope can have the same mass-radius relation as a planet containing ices, but no H/He. 

A large portion of the exoplanets discovered so far are close-in planets  with a semi-major axis $<$ 0.1 AU. At such small distances from the host star, planets are exposed to  strong stellar irradiation and can undergo (hydrodynamic) atmospheric evaporation, a process tha{t} can be observed \citep{Vidal-Madjar2003,Ehrenreich2015}. Low-mass low-density planets at close-in orbits are most likely to lose their entire gaseous envelopes due to their small gravitational binding energy \citep[e.g.,][]{Lammer2009,Lopez2012,Owen2012,Jin2014}, and consequently become bare solid planetary cores with larger mean densities. 

{A gaseous H/He envelope will lead to a significant jump in planetary size due to the low density of gas. Additionally, the timescale of losing the last radius-increasing H/He is short  ($\sim10^{5}$ yr) compared to typical  ages of planets ($10^{9}$ yr). Thus, the stripped bare planetary cores will be clearly separated from the planets that still retain an H/He envelope in planetary sizes, which results in an underpopulated gap or valley in the planetary radius distribution \citep{Owen2013,Lopez2013,Jin2014,LopezRice2016,ChenRogers2016}. This feature can be used as a powerful tool to study the low-mass close-in planets {\citep{Lopez2013}}. For example, {where the gap is located}, and how {it} change{s} towards longer semi-major axes can be a criterion of their {bulk composition and} formation history, {distinguishing} in situ formed rocky planets {from} stripped cores of {icy} sub-Neptunes {as well as different envelope loss processes} \citep{Lopez2013, LopezRice2016}.

For individual planets,}  the bare cores have a smaller compositional degeneracy because the extra degeneracy introduced by a gaseous H/He envelope is removed. Thus,  with sufficient accuracy in the measurement of the mass and radius of close-in low-mass planets, we can better constrain the bulk composition by assuming that they are bare cores, in particular regarding the question whether some of these low-mass cores contain  large amounts of ices. For a planet core that was fully formed outside of the (water) iceline, an ice mass fraction of about 50\% is expected from condensation models. The addition of other ice species like CO$_{2}$, CH$_{4}$ or NH$_{3}$ could increase the ice mass fraction even to about 2/3 \citep{Lodders2003,Min2011}. In this work, we are therefore mainly interested in the statistical population-wide consequences of the presence of large amounts of ices that substantially alter the mass-radius relation relative to a purely rocky composition. Such a characterization in terms of a large ice fraction is as mentioned of high interest for formation and evolution models. For the detailed analysis of individual objects, the reader is referred to, e.g., \citet[][]{Dorn2017a,Dorn2017b}.

In this work, we investigate the population-wide impact of atmospheric escape of the primordial H/He envelope on two different synthetic planet populations, one with rocky planetary cores and the other with icy cores. We find that the final observable properties of close-in planets depend on the bulk composition of the planetary cores. For example, the typical statistical population-wide imprints of evaporation, like the locus of the ``evaporation valley" or  the one-dimensional bimodal radius distribution  differ depending on the core composition. They may also be erased or blurred if there are both rocky and icy cores at close-in orbits {\citep{Lopez2013}}. 

We furthermore find that the planetary mass vs. mean density (mass--density) diagram of a planet population reveals important features of planet formation and evolution \citep{Rauer2013,HatzesRauer2015,Baruteau2016}. It allows in particular to identify more clearly than in the mass--radius diagram the different fundamental planetary compositions: solid planets with rocky or icy interiors (terrestrial and ocean planets); core dominated planets with low amounts of H/He that did not trigger gas runaway accretion like (sub-)Neptunian planets; and gas giant planets dominated by H/He that did trigger it (Jovian planets). Observing these transitions is also of high interest for formation models, as it makes it possible to better understand the governing physics of planet formation. Mechanisms that can be constrained in this way are for example the (grain) opacity in primordial atmospheres and the associated efficiency of H/He accretion \citep{Podolak2003,Ormel2014,Mordasini2014}, envelope enrichment \citep{Venturini2016} or  the hydrodynamics of embedded primordial atmospheres \citep{Ormel2015}. They influence the amount of H/He that can be accreted  by a planet and thus the mean planetary density as well as the critical core mass when gas runaway accretion can start. We find, however, that it is necessary to take into account the subsequent evaporation during the evolutionary phase as evaporation can substantially reduce the H/He mass compared to the post-formation one, at least for planets inside of $\sim1$ AU (depending on mass).

Moreover, thanks to the evolution of the mass--density diagram in time it may be possible to remove or at least reduce the degeneracy in the compositional parameters of close-in low-mass planets. For this it is important to consider that the mass--density relation is not static in time depending on the planet type, but evolves because of contraction and evaporation. This means that statistically solid and gaseous can be distinguished by studying the mass--density relation for a given mass and distance (or irradiation) interval at different moments in time, like for example 100 Myr and 5 Gyr. With the exception of extremely close-in very low-mass planets \citep{perez2013}, for solids planets, the density is nearly constant in time, while for planets with significant gaseous envelopes it increases in time. Statistically, this would mainfest in e.g. a increase of the mean density for the considered sub-population, allowing to probe the typical composition in various parts of the mass-flux space. {For low-mass gas-poor planets, different ice fractions will lead to different evolution tracks due to the changes in heat capacities and density distributions, and this can be used to statistically constrain the volatile content of planets if their radius and ages can be measured with sufficient precision \citep{Alibert2016}}. To date, no precise age determinations for a statistically large sample of transiting planets/host stars on the main sequence have been possible. But with the PLATO 2.0 satellite \citep{Rauer2013} scheduled for launch in  2024, the ages of a high number of host stars will be determined with about 10\% accuracy thanks to systematic astroseismological analyses. This will enable for the first time to observationally follow the evolution of the planetary population in time, putting  constraints not only on evaporation models like presented here, but also on models for inflated planets \citep[e.g.,][]{Batygin2011} or the (re)distribution of heavy elements \citep[e.g.,][]{Vazan2013}.

The paper is organized as follows. We briefly describe our planet evolution model with evaporation in Section \ref{model}.  In Section \ref{rockyicy}, we show the 2D radius-distance distribution and the associated 1D radius distribution of the synthetic rocky and icy core populations and compare with observations. In Section \ref{sect:icemassfraction} we analyze the ice mass fraction of planets that should have lost their primordial H/He. In Section \ref{massdensity}, we study the mass--density distributions of the rocky and the icy core populations as a function of time and distance, and compare our results with the observed mass--density distribution of exoplanets. We finally present a brief summary and  discussion in Section \ref{discussion}. 

\section{Model for Planetary Thermal Evolution and Evaporation}\label{model}
Our model of combined thermodynamical evolution and atmospheric escape was described in detail elsewhere \citep{Mordasini2012a,Jin2014}, therefore we give here only  a short summary. We reiterate just two aspects of the evolution model, which is first the outer boundary condition and second the rate of atmospheric escape. In all calculations, we set the start of the planetary evolution stage at 10 Myr, because nearly all protoplanetary gas disks have disappeared at about 8 Myr  in our planetary formation model \citep{Alibert2005,Mordasini2012a,Mordasini2012b,Mordasini2014ija}. From then on, the disk-driven migration and accretion of mass (planetesimals and nebular gas) stops, and the planets enter the evolutionary stage.  

\subsection{{Atmospheric structure model}}
One important aspect in modeling the long-term thermodynamical evolution of close-in planets is the temperature profile used to calculate the atmospheric structures in the top of the atmosphere where the stellar irradiation can penetrate.
Analytical temperature profiles were developed based on the two-stream approximation, which assumes that there is an incoming stellar irradiation flux in the visible wavelength range and an absorbed and intrinsic thermal flux \citep{Hubeny2003,Hansen2008,Guillot2010,Heng2012,Robinson2012,Parmentier2014,Heng2014}.
In this work, we adopt the globally averaged temperature profile from \citet{Guillot2010} ($\tau$ is the optical depth):
\begin{equation}
  \begin{split}
    T^4 &={3\tint^4\over 4}\left\{{2\over 3}+\tau\right\}+ {3\teq^4\over 4}\left\{{2\over 3}+\right.\\
    & \quad \left. {2\over 3\gamma}\left[1+\left({\gamma\tau\over 2}-1\right)e^{-\gamma\tau}\right]+ {2\gamma\over 3}\left(1-{\tau^2\over 2}\right)E_2(\gamma\tau)\right\}
  \end{split}
 \label{2bdglobal}
\end{equation}
where $\tint$ is the intrinsic temperature that characterizes the heat flux from the planet's interior,
$\teq$ is the equilibrium temperature obtained by averaging the stellar radiation over the entire planet surface, and
$\gamma=\kappa_{\rm v}/\kappa_{\rm th}$ is the ratio of the visible opacity to the thermal opacity \citep{Guillot2010}.
The visible opacity $\kappa_{\rm v}$ is not explicitly calculated but is incorporated in the model by $\gamma$, which was tabulated in \citet{Jin2014}. $E_2$ is the exponential integral $E_n(z)\equiv\int_1^\infty t^{-n}e^{-zt}dt$ with $n=2$.

\subsection{{Evaporation model}}
Our main focus in the evolution model is the atmospheric escape due to the heating from stellar X-ray and extreme-ultraviolet (XUV) irradiation \citep[e.g.,][]{Watson1981,Lammer2003,Baraffe2004,Yelle2004,Tian2005,Murray-Clay2009,Owen2012,Owen2016}. Depending on the locations of the ionization front created by EUV flux and the sonic point in X-ray-driven flow \citep{Owen2012}, an escaping wind can be dominated by either X-ray or EUV heating. Typically, atmospheric escape is in the X-ray-driven regime during the early evolution stage \citep{Owen2012,Jin2014}. 

To describe this regime we use the energy-limited escape rate given by \citet{Jackson2012} with the X-ray flux from 1 to 20 ${\rm \AA}$ taken from \citet{Ribas2005}. The typical values of the heating efficiency $\epsilon$ in energy-limited model are in the range of 0.1--0.25 \citep{Lammer2009,Jackson2012}. We set $\epsilon$ in the X-ray-driven regime to 0.1, considering that mainly the X-ray flux from 5 to 10 ${\rm \AA}$ is responsible for heating \citep{Owen2012}, rather than the X-ray flux from 1 to 20 ${\rm \AA}$.

After the early evolution phase of intense X-ray-driven evaporation, atmospheric escape will transition to the EUV-driven regime, which itself can be further divided into two sub-regimes \citep{Murray-Clay2009}. Here we adopt the temporal evolution of the EUV luminosity of a solar-like star from \citet{Ribas2005}. The significant spread in the XUV luminosity among different stars of similar spectral type because of different rotation rates \citep{Tu2015} is  not yet taken into account in the simulations presented here.

At high EUV fluxes, a large portion of the heating energy is lost to cooling radiation, thus the energy-limited approximation is not suitable anymore. In this case we use the radiation-recombination-limited approximation given by \citet{Murray-Clay2009}:
\begin{equation}
  \dot{M}_{\rm rr-lim} \sim 4 \pi \rho_{\rm s} c_{\rm s} r_{\rm s}^2
\end{equation}
where $c_{\rm s}$ is the isothermal sound speed, $\rho_{\rm s}$ the gas density at the sonic point, and $r_{\rm s}$ the radius where the escaping flow reaches the sonic point. These quantities can be estimated using the description of \citet{Murray-Clay2009}. At low EUV fluxes ($<10^{4}$ erg cm$^{-2}$ s$^{-1}$), the mass-loss rates can again be estimated using again the energy-limited approximation \citep[e.g.,][]{Watson1981,Murray-Clay2009}:
\begin{equation}
  \dot{M}_{\rm e-lim}=\epsilon\frac{\displaystyle{\pi F_{\rm EUV} R^3_{\rm base}}}{\displaystyle{G M_{\rm p} }}
  \label{Mdotxray}
\end{equation}
where $\epsilon$ is the heating efficiency, $F_{\rm EUV}$ the EUV flux at the position of the planet, $R_{\rm base}$ the radius of the photoionization base \citep[estimated as in][]{Murray-Clay2009}, $M_{\rm p}$ the planet mass and $G$ the gravitational constant.
 Here we adopt the heating efficiency found in \citet{Murray-Clay2009}, $\epsilon=0.3$. 
 
Note that in reality, the regime of atmospheric escape for a specific planet, and the heating efficiencies in each regime  would depend on the specific planetary mass, radius, atmospheric composition, and the stellar flux and would thus change with time \citep{Yelle2004,Tian2005,Owen2012}. Detailed criteria for the occurrence range of different regimes were recently given in \citet{Owen2016}; they will be included the population synthesis model used here in future work. On the other hand, the statistical imprints of evaporation on the entire planet population do not vary dramatically when the intensity of evaporation is varied within a reasonable range as shown in \citet{Jin2014}. 

\begin{figure}
 \includegraphics[width=8.9cm]{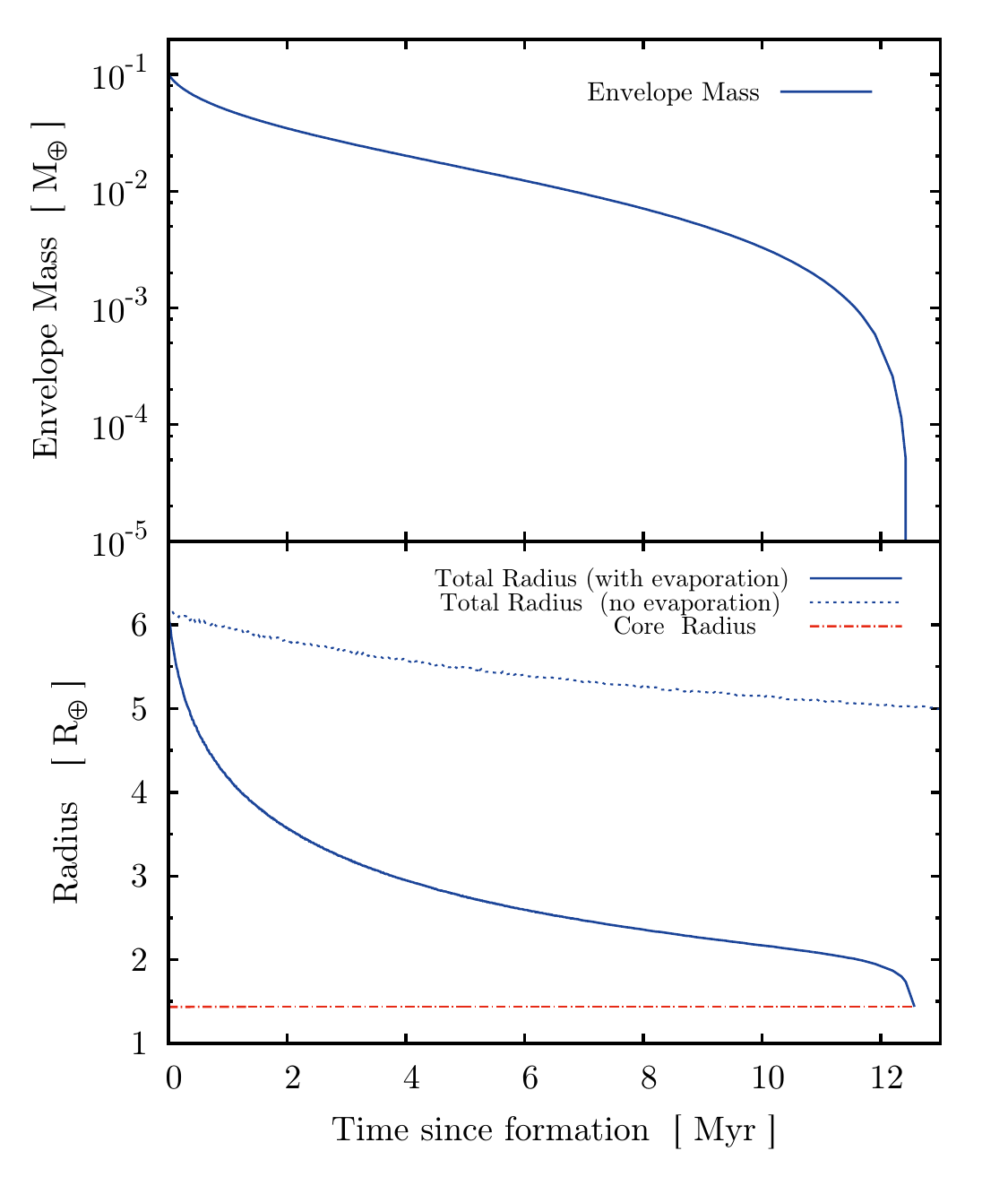}
 \caption{Temporal evolution of the envelope mass and total radius of a close-in low-mass planet. This planet has a rocky core of 4 $M_{\oplus}$  and an initial  H/He envelope of 0.1 $M_{\oplus}$. The blue solid lines show the simulation that includes evaporation. The blue dotted line shows the same simulation but  without evaporation. The red dash-dotted line shows the core radius. There is a substantial decrease in planetary radius on a short timescale of $\sim10^{5}$ yrs when the planet loses the last part of its H/He envelope.} \label{single}
\end{figure} 

Figure \ref{single} shows as an example the temporal evolution of a 4 $\mearth$ planet at an orbital distance of 0.05 AU. Such low-mass planets can have a large radius even with a tenuous envelope \citep{Adams2008,Rogers2011,Mordasini2012b}, and can be easily evaporated to bare cores at close-in orbits \citep{Lopez2013,Jin2014}.  The planet shown in Figure \ref{single} has an Earth-like core\footnote{In this publication we follow the astrophysical nomenclature of calling the entire solid part of the planet consisting of iron, silicates, and potentially ices ``core''. This is different from the geophysical meaning.} (2:1 silicate:iron mass ratio) of 4 $M_{\oplus}$ and a primordial H/He envelope of initially 0.1 $M_{\oplus}$. Its initial luminosity is 0.1 $L_{\rm Jupiter}$, corresponding to a specific entropy of 7.5 $k_{\rm b}/{\rm baryon}$ at the core-envelope boundary. If atmospheric escape is included, this planet loses all its H/He envelope already at $\sim$ 12.6 Myr after the start of evolution. Compared to the simulation that does not include atmospheric escape, the planetary radius decreases significantly in the case evaporation is included. The importance of already a small amount of H/He on the total radius becomes especially clear in the later stage when the planet is losing its last $\sim10^{-3}$ $M_{\oplus}$ of envelope, during which the planetary radius decrease rapidly on a timescale of $\sim10^{5}$ years from $\sim$ 1.8 $R_{\oplus}$ to $\sim$ 1.4 $R_{\oplus}$ (the radius of its bare rocky core).

\subsection{{Limitations of the model}}\label{subsect:limitationsmodel}
We recall a number of limitations of our model that should be critically kept in mind especially regarding the comparisons with observations: first, as mentioned, in reality there is a wide intrinsic spread in stellar $L_{\rm XUV}$ of almost two orders of magnitude at early ages when evaporation is most important \citep{Tu2015}. It is however clear that this spread in $L_{\rm XUV}$, as well as different efficiencies $\epsilon$ of evaporation (here also fixed to one value) could have important effects on the impact of evaporation \citep{Tu2015}. {These variations affect the population-wide impact of evaporation and} could make the imprints of evaporation fuzzier, potentially more similar to the observational data. {This is the subject of a follow-up paper.}

Second, in the model all planets start with a primordial H/He envelope and reach their final mass during the presence of the protoplanetary gas disk. In reality, especially low-mass planets can acquire their final mass  only well after the dissipation of the gas disk \citep[e.g.,][]{Baruteau2016}, such that they start without H/He envelopes which would also weaken the evaporation imprints. Finally, and related to the second point, (giant) impacts can also play an important role in removing envelopes, leaving different statistical imprints \citep[e.g.][]{Schlichting2015,LopezRice2016}. {Moreover, besides impacts there could be additional loss mechanisms like for example mass loss driven by magnetohydrodynamic waves \citep{Tanaka2014} or core-powered mass loss \citep[][see also \citealt{Owen2016b}]{ginzburg2016,ginzburg2017}.} A combination of all these effects would then lead to the observed radius-distance (or flux) distribution. 

\section[]{The locus of the evaporation valley for rocky and icy cores}\label{rockyicy}
We simulate the long-term evolution of two planetary populations with rocky or icy cores, respectively, using the planetary population synthesis models that include both planet formation \citep{Alibert2005}, as well as the subsequent long-term evolution (cooling and contraction) and atmospheric escape \citep{Mordasini2012a,Jin2014}. 

\begin{figure*}
\begin{center}
 \includegraphics[width=\textwidth]{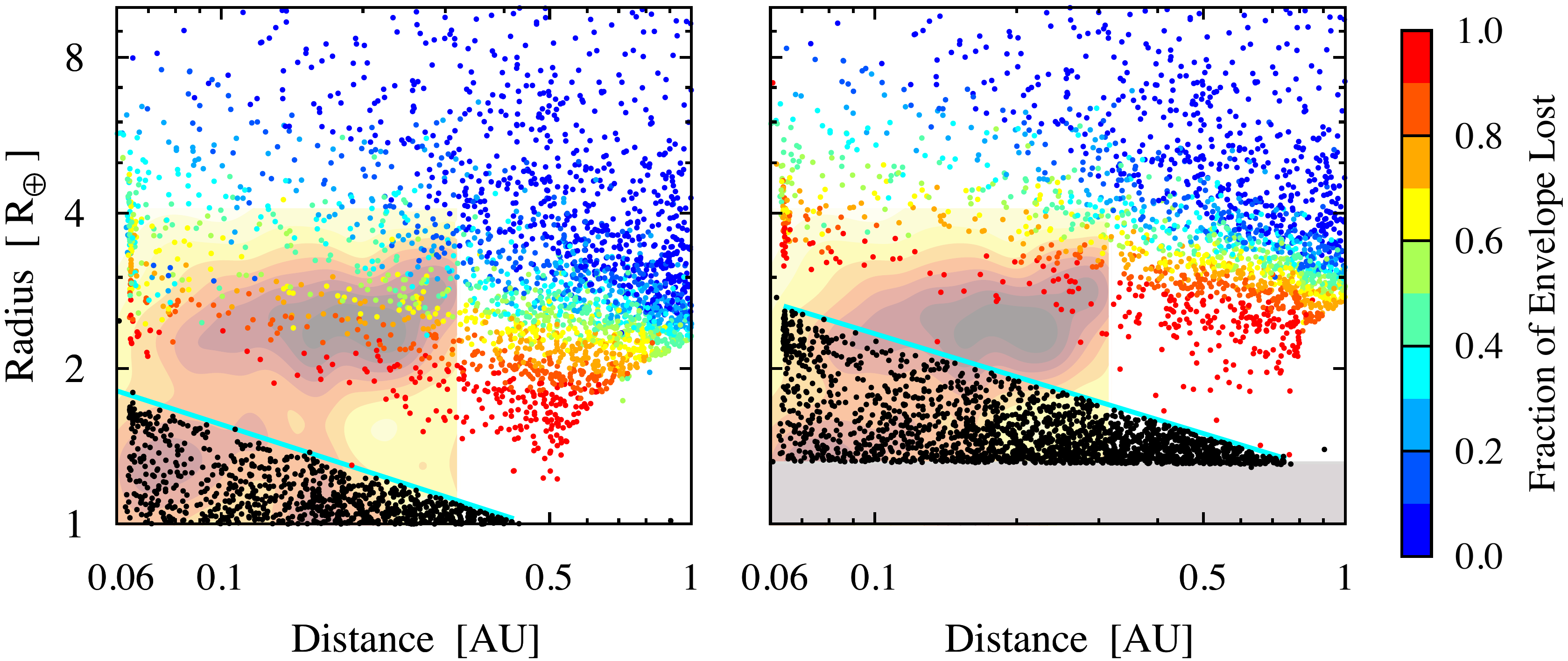}
 \caption{$a$--$R$ distributions of the two synthetic planetary populations at 5 Gyr and comparison with observations. The points in the left panel show the rocky core population in which all planets have rocky cores (2:1 silicate:iron ratio), while the right panel shows the icy core population with icy cores (75\%  ice  in mass). The color of each point shows the fraction of the initial envelope that was evaporated. Black points are the planets in the ``triangle of evaporation'' that have lost all their  H/He. In the rocky population, the ``evaporation valley" occurs at $\sim$1--2$\rearth$ depending on distance. The cyan line showing the largest bare core is here at $R_{\rm bare}\approx1.6\times (a/0.1\ $AU)$^{-0.27} \rearth$. In the icy core population, the valley is at $\sim$ 2--3 $R_{\oplus}$. The cyan line is here at $R_{\rm bare}\approx2.3\times (a/0.1 \ $AU)$^{-0.27} \rearth$. The gray shaded region in the right panel remains empty because only planets more massive than 1 $M_{\oplus}$ are included. The empty arc-like part in the  bottom right corner is also an artifact of this minimal mass and has no physical meaning. The contours are from \citet{Fulton2017} and show the completeness corrected occurrence rate of Kepler planets with brown (yellow) indicating a high (low) occurrence. The observed location of the valley is compatible with a predominantly rocky core composition (left), but inconsistent with a mainly icy composition (right).}
  \label{twopop}
  \end{center}
\end{figure*}

During formation, the disk-driven migration of each planet was calculated using the isothermal Type I migration rate \citep{Tanaka2002} with a reduction factor of 0.1, in combination with the evolution of the protoplanetary gas disk \citep{Alibert2005,Mordasini2012a,Mordasini2012b}. Since we want to focus on the differential impact of the core composition on a planetary population undergoing atmospheric mass loss, we use the one-embryo-per-disk model. Planet-planet scattering is thus not included; the effect of the concurrent formation of several planets can be found in \citet{Alibert2013}.

The only difference between our two synthetic populations is the composition of planetary cores, that we artificially impose in order to study the limiting cases. In the rocky population, all the planetary cores have an identical entirely rocky composition with a 2:1 silicate:iron ratio, as in Earth. In the icy population, 75\% of the core mass is in contrast water ice, and the other 25\% is the same rocky material that has a 2:1 silicate:iron ratio. This high water mass fraction that is inspired by the original \citet{Hayashi1981} minimum mass solar nebula model is chosen to make the difference between the two populations most apparent. In reality, one would expect a range of ice mass contents, depending on the exact formation location and chemical composition of the disk \citep[e.g.,][]{Mordasini2016}.

Planets that have such a large amount of ice in their cores can only be formed beyond the snow line. These icy planetary cores could then migrate to close-in orbits by disk-based migration \citep{Goldreich1980,Lin1996,Zhou2005} or migration due to the dynamics in the planetesimal disk \citep{Terquem2007,Ji2011,Ormel2012}. Despite the fact that the link is not self-consistent in the current work as we artificially set the core composition for all planets, it is clear that the rocky population can be associated with a formation of close-in low-mass planets inside of the iceline, while the icy population represents the case of efficient inward migration from beyond the iceline \citep{Baruteau2016}. These cases represent limiting cases of the different theories for the formation of the numerous class of close-in, low-mass planets \citep[e.g., ][]{idalin2010,chianglaughlin2013,raymond2014,Baruteau2016}.

The mass--radius relationship of known exoplanets show that a number of them are at least consistent with models of a high water content and no significant envelopes \citep{Howe2014} (but see also \citealt{Lopez2016}). It is clear that in reality, not all the close-in planets of an actual population will have such a large amount of ice content in their cores, but using an entirely icy population is helpful to make clear the population-wide impact of the bulk composition of planetary cores.

\subsection{{The locus in the $a$--$R$ plane}}
We evolve planets with a mass of at least 1 $\mearth$ in the rocky and icy core populations for 10 Gyr with atmospheric escape included. The population-wide impacts of evaporation and how they are related to the parameters of the evaporation model have been extensively studied in \citet{Jin2014}. Here we focus on the influence of the bulk composition of planetary cores, but recall the following:
According to the core-accretion paradigm, low-mass cores can only accrete a small amount of gas due to their long Kelvin-Helmholtz timescales. Their initial envelopes, typically a few percent of the total planetary masses, can be entirely evaporated in a relatively short timescale for planets that have a sufficiently low (core) mass and small orbital distance, as illustrated by the example in Figure \ref{single}. The radius of a bare core is substantially smaller than the radius of a planet that has a gaseous envelope. Moreover, the loss of the last 0.1\% of a planetary envelope occurs on a timescale of $\sim$ $10^{5}$ yrs (Figure \ref{single}), {much less than the typical age of planets ($\sim$ $10^{9}$ yrs)}, therefore it is unlikely to see a planet exactly in this period. As a result, an ``evaporation valley" running diagonally downward appears in the semi-major axis vs. radius distribution, corresponding a region that is devoid of planets, after most of the low-mass planets become bare cores \citep{Owen2013,Lopez2013,Jin2014,LopezRice2016,ChenRogers2016}.

Figure \ref{twopop} compares the $a$--$R$ distribution at 5 Gyr of the rocky and icy core populations. Both populations show an evaporation valley of $\sim$ 0.5 $R_{\oplus}$ in width. But the locations of the evaporation valley in these two populations are  clearly different. In the rocky population, the valley occurs at $\sim$ 1--2.3 $R_{\oplus}$ depending on distance, whereas in the icy population, the evaporation valley occurs at $\sim$ 1.3--3 $R_{\oplus}$. The two cyan lines in Fig. \ref{twopop} at the bottom of the valley showing the largest bare cores (solid planets without H/He) as a function of distance are at 
\begin{equation}\label{eq:Rbarrocky}
R_{\rm bare,rocky}\approx1.6\times (a/0.1 \ \mathrm{AU})^{-0.27} \rearth
\end{equation}
for the rocky core population. This is consistent with the transition found by \citet{LopezRice2016}. In the icy core population, this limit is at  
\begin{equation}
R_{\rm bare, icy}\approx 2.3\times (a/0.1 \ \mathrm{AU})^{-0.27} \rearth.
\end{equation}
The middle of the gap lies about 0.3 $\rearth$ above these values. There are two reasons for the different location of the valley for rocky and icy cores: first, in the icy population, when the envelope is still present, the mean density is lower because of the icy cores (provided that the envelope mass fraction is $\lesssim0.1$, \citealt{Mordasini2012b}). This makes the planets more vulnerable to evaporation, shifting the limit not only to larger radii, but also higher masses (see below). Second, once the envelope is lost, the sizes of the bare cores in the icy population at fixed mass are substantially larger due to the 75\% ice content. 

The region in the $a$--$R$ plane of planets having lost all H/He (the black dots in Fig. \ref{twopop}) below the cyan line has in a log-log plot a triangular shape. Therefore, we call this region the ``triangle of evaporation''. We further study the composition of planets in this interesting region in Sect. \ref{sect:icemassfraction}. {One notes that the upper boundary of the triangle of evaporation is very sharp for the two synthetic populations. This is partially an artifact of the following two aforementioned model simplifications: First, all planet cores have the same Earth-like silicate:iron ratio, and, for the icy population, ice content. Second, the stellar XUV luminosity as a function of time is identical. In reality, there are variations in these quantities  making the transition fuzzier, potentially as it is seen in the observational data (see the discussions in Sect. \ref{subsect:limitationsmodel} and \ref{subsect:ironmassfraction}).}

While there is a partial overlap in the location of the valleys {in the rocky and icy core populations}, we also see that a large number of the bare icy cores are of the sizes that correspond to the evaporation valley in the rocky population. Therefore, if  close-in low-mass planets in a population consist of both rocky and icy cores in appropriate ratios, there will be a less clear evaporation valley after low-mass planets have lost all their envelopes. As already noted by \citet{Lopez2013}, the presence or absence of the evaporation valley as well as its location and depth can thus serve as a test whether close-in low-mass form with or without large quantities of water, a crucial information to understand their formation. 

\subsection{{The situation in the $a$--$M$ plane}}
We note that in contrast to the radius-distance distribution, the mass-distance distribution of low-mass planets does not contain a gap or valley at least outside of 0.06 AU, as the envelope masses are negligible compared to the total planetary mass, such that their loss does not affect in a significant way the $a-M$ diagram, as shown in \citet{Jin2014}. For the type of planet considered here, the mass distribution therefore reflects the formation, while the radius distribution nowadays is driven by evolution. Interestingly, the locus of the gap which is a consequence of evolution allows to constrain their formation (inside vs. outside the iceline) better than without such an evolutionary effect. 

The transition masses corresponding to the cyan lines at the upper boundary of the triangle of evaporation in Fig. \ref{twopop} are at about 
\begin{equation}
M_{\rm bare,rocky}\approx 6\times (a/0.1 \ \mathrm{AU})^{-1} \mearth
\end{equation}
for a rocky core composition and   
\begin{equation}
M_{\rm bare,icy}\approx 8\times (a/0.1 \ \mathrm{AU})^{-1} \mearth
\end{equation}
in the icy core population. These radius and mass limits allow to identify different planet types, as demonstrated below  in Sect. \ref{sect:icemassfraction}. 

We comment that another evaporative desert in the $a$-$M$ diagram that is not related to the low-mass and small planets discussed here may also exist. It is relevant for planets with larger initial masses (very close-in, not very massive giant planets), where the loss of the envelope does lead to a significant reduction of the total mass, in contrast to the low-mass planets we consider here. Indeed, there is a desert in the observed mass-distance diagram centered at about 60 $\mearth$ and $a\approx0.03$ AU  \citep[e.g.,][]{kurokawa2014,mazeh2016}.

\subsection{Comparison with Kepler observations}
The brown-yellow contours in Fig. \ref{twopop} show the completeness-corrected relative occurrence rates of Kepler planets derived by \citet{Fulton2017}. The observational data also contains a valley at about 1.7 $\rearth$, separating a super-Earth local occurrence maximum of smaller, closer-in planets from a sub-Neptune local occurrence maximum of larger, more distant planets. One sees that the location of the observed valley is compatible with the synthetic rocky core population, but not with the icy core population. In the latter, the observed  occurrence maximum of  sub-Neptune planets  would fall into the predicted valley. This shows that a predominately icy core composition is inconsistent with observations. In the rocky core  population, the location of both the super-Earth and sub-Neptune over-densities are in contrast similar to the observations. We note that the observations can currently not constrain the radial dependency of the transition because of a lack of completeness of small planets at larger distances \citep{Fulton2017}. Probing such a region will be an important task for future transit observations, allowing to disentangle the different mechanism{s} that lead to bare cores. 

If the gap is really due to atmospheric escape, then we can conclude from this comparison that the cores of close-in low-mass Kepler planets are predominantly composed of silicates and iron, without large amounts of ices.  This is the most important result of this paper. The same conclusion was recently reached by \citet{Lopez2016} from an analysis of a different aspect, namely the radii of ultra-short-period planets. The location of the valley in the rocky population is also compatible with the transition to non-rocky planets at about 1.6 $\rearth$ found by \citet{Rogers2015}. This suggests that these planets have accreted mainly inside of the water iceline. Combined with the clear population-wide imprints of past orbital migration in the Kepler data like in particular the frequency maxima just outside of MMR period ratios \citep{Fabrycky2014}, the global picture arises that orbital migration in the protoplanetary disk played an important role in the formation of these planets, but that the migration was confined to the part of the disk inside of the water iceline. A reason for this separation could for example be Type I migration traps that occur at opacity transition like the water iceline \citep[e.g.,][]{dittkristmordasini2014}, and the simple effect that lower-mass planets migrate slower than more massive ones in Type I migration \citep[e.g.,][]{Ward1997}. 

These effects could mean that low-mass planets with masses of $\sim$5 $\mearth$ or less forming outside of the water iceline did not have time to migrate all the way to 0.1 AU or were stuck in the migration traps. More massive (sub-)Neptunian planets with masses of $\sim$10 $\mearth$ or more could in contrast still have migrated from beyond the iceline to 0.1 AU or less, because of their faster migration rate and the saturation of the positive (outward) corrotation torque at higher masses causing them to leave the migration traps. This would mean that there could be an ice mass fraction that increases with mass among the close-in planets. We note that population syntheses including these effects \citep{Alibert2013} indeed predict such a "vertical" (in the $a$--$R$ diagram) compositional gradient to increasing ice mass fraction with increasing mass or radius for planets inside of about 0.5 AU. Unfortunately, this class of planets is too massive to be probed with the position of the evaporation valley except for very close planets (Sect. \ref{sect:icemassfraction}). The fact that these (sub-)Neptune planets usually can retain thick H/He envelopes makes it very difficult to derive their core composition from the density. 

\subsection{The bimodal radius distribution}
It is interesting to see how the two-dimensional distance-radius distributions translate into the one-dimensional radius distributions, and to compare them with Kepler observations. Since for evaporation the transition is a function of distance, in this marginalization the distribution of orbital distances also matters. We first compare to the older data from \citet{Petigura2013} and then to the more recent \citet{Fulton2017} analysis. 

\begin{figure}
\includegraphics[width=8.9cm]{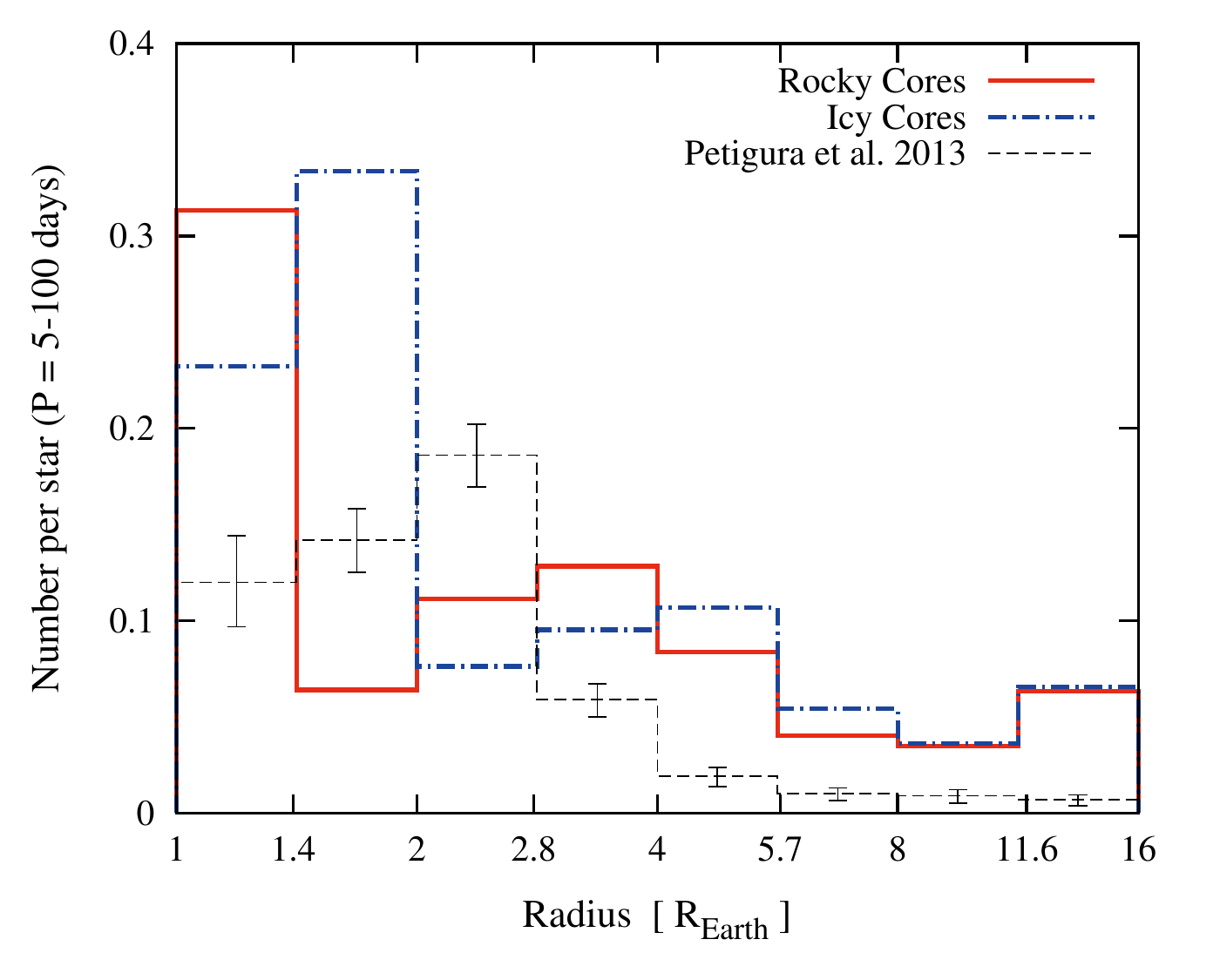}
 \caption{Histogram of radii of close-in planets with orbital period between 5 and 100 days in the synthetic rocky (red solid) and  icy core (blue dashed-dotted line) populations. The black dashed line with error bars shows the occurrence of  Kepler candidates with a correction for survey completeness from \citet{Petigura2013}. The bimodal size distribution at small planetary sizes would be removed or reduced if both rocky and icy cores existed at close-in orbits.}
\label{compkepler}
\end{figure}

Figure \ref{compkepler} shows the occurrence rate as a function of radius for planets with orbital periods between 5 and 100 days in the two synthetic planet populations. For comparison, the plot also shows the occurrence rate of the Kepler candidates with a correction for survey completeness from \citet{Petigura2013} in the same rather wide bins. The sizes of the bare low-mass cores in the rocky population are in the range of about $\sim$ 1--1.9 $R_{\oplus}$, while in the icy population, the bare low-mass cores are in the range of 1.3--2.5 $R_{\oplus}$.

In Figure \ref{compkepler}, a depletion of planets (a local minimum in the radius histogram) is seen in the bin at 1.4--2 $R_{\oplus}$ in the rocky population, and in the bin at 2--2.8 $R_{\oplus}$ in the icy population. The maximum in the radius distribution of the bare cores in the icy population occurs at the same locus as the minimum in the rocky core population. Therefore, the local minimum in the one-dimensional (bimodal) radius distribution at small planetary sizes \citep{Owen2013,Jin2014} would be reduced by an appropriate combination of rocky and icy cores for such wide bins. Large uncertainties in the radius measurements would have a similar effect. In other words, if an important part of close-in exoplanets would have a core with a large ice mass fraction, there would be no obvious observational imprint (bimodal size distribution, evaporation valley) caused by atmospheric escape in observational data. 

Comparison with the observed distribution of radii in Fig. \ref{compkepler} from \citet{Petigura2013}  shows that there is indeed no local minimum in this early set of observational analysis. This would lead to the conclusion that the cores have a mixed icy and rocky composition (we will see next that this is not the case). On the other hand, with finer bins, \citet{Owen2013} had found a rather shallow local minimum at about 1.9 $\rearth$ among the Kepler KOIs. At the time of writing this work, the radius distribution of confirmed Kepler candidates at the NASA exoplanet archive still shows such a local minimum at around 1.7 to 1.9 Earth radii. 

\begin{figure*}
 \includegraphics[width=18cm]{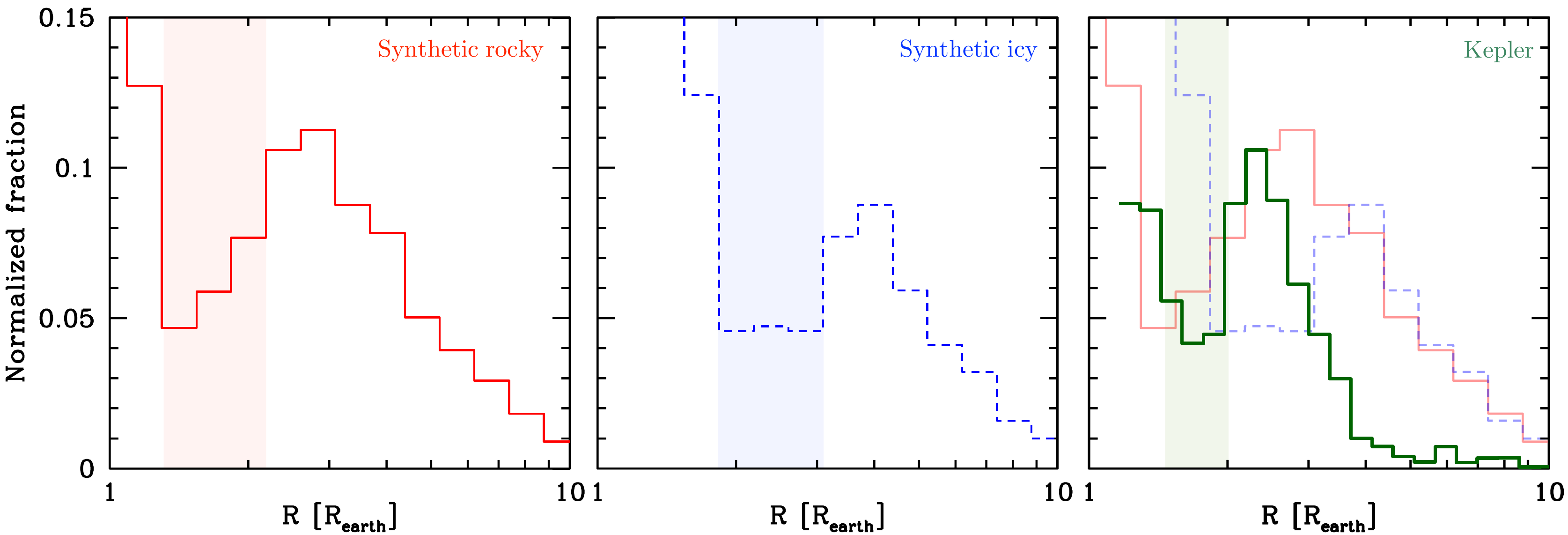}
 \caption{Comparison of the radius distribution in the two synthetic populations and in the completeness-corrected Kepler distribution of \citet{Fulton2017}. The left and center panels show the synthetic population with rocky (red) and icy cores (blue dashed) separately. The position of the evaporation valley is shaded. The observed distribution (right panel, green) was normalized in the bin in the second maximum (the sub-Neptune maximum at about 2.3 $R_\oplus$) to have the same value as the red curve at this point. We see that the location of the minimum in the rocky core population (red shaded, 1.3-2.1 $R_\oplus$) is consistent with the observations, whereas the minimum in the icy core population (blue shaded, 1.8-3  $R_\oplus$) is at too large radii. In the icy core population the minimum occurs at the position of the observed maximum, showing that a mainly icy core composition is inconsistent with observations. }
  \label{compkeplernew}
\end{figure*}

Recently, \citet{Fulton2017} presented a new analysis of the radius distribution of small Kepler planets. Their new detailed spectroscopic characterization of the host stars reduces the median uncertainties in the stellar properties like radius and thus also the planetary radius, enabling them to see finer structures. 

They found a clear bimodal distribution that is shown by the green line in the  right panel of Fig. \ref{compkeplernew}. The distribution consists in order or increasing radius of a first local maximum at a radius of about 1.3 $\rearth$ (the super-Earth maximum). It is followed by a deep local minimum centered around about 1.7 $\rearth$. This gap is about 0.5 $\rearth$ wide. The second local maximum (the sub-Neptune maximum) follows at around 2.3 $\rearth$. The decrease in frequency in the gap relative to the two approximately equally high surrounding maxima is approximately a factor 2-2.5, i.e., much larger then in the \citet{Owen2013} analysis. At even larger radii, beyond the sub-Neptune maximum, the well-known decrease in planet frequency with increasing $R$ follows \citep[e.g.,][]{Borucki2011}. Interestingly, this newly observed structure is quite comparable to the synthetic radius distribution predicted theoretically in \citet[][their Fig. 14]{Jin2014}, where a strong imprint of evaporation with a deep evaporation valley in the radius distribution was found in the models. At the time of writing of  \citet{Jin2014}, this appeared rather inconsistent with the Kepler data available at that time.

In the left and center panel of the plot, we show the radius distribution of the rocky and icy core populations in the same finer bins. In order to have a sufficient number of synthetic planets, we  include synthetic planets with a semimajor axis of less than 0.6 AU, about 0.2 AU more than in the \citet{Fulton2017}. The comparison of the two synthetic populations with the observed distribution reveals several interesting matches but also differences. 

(1) It shows in a more precise way (compared to Fig. \ref{compkepler}) the location of the minima in the two synthetic populations. We see that the minimum in the rocky core population is centered at 1.3-2.1 $R_\oplus$, whereas in the icy core it is  at 1.8-3  $R_\oplus$. These radius intervals are shaded in red and blue in the figure.  There is a slight overlap of the two theoretically predicted gaps at around 2 $R_\oplus$.

(2)  Most importantly, the comparison with the observed 1D distribution shows clearly that the position of the valley  in the synthetic rocky core population is consistent with the \citet{Fulton2017} observations whereas the position of the valley in the synthetic icy core composition is, in contrast, inconsistent. The minimum occurs for these icy core compositions at too large radii, in a way that the theoretically predicted minimum occurs quite exactly where in the observations there is a maximum (the sub-Neptune maximum).  This finding that the 1D distribution for a predominately rocky composition is consistent with observations, but not for an icy  one, reflects the equivalent findings for the 2D distance (or flux)-radius distributions presented above (Fig. \ref{twopop}). Together they form the main result of this study.

(3) A detailed comparison of the gap shape in the synthetic rocky and observed population shows that the decrease into the gap coming from the right (from large radii) agrees rather well.  The largest difference between the rocky and actual population only occurs at radii between 1.2 to 1.6 $\rearth$, where the synthetic population is still strongly depleted, but where the observations already show the super-Earth peak. These planets could be massive rocky planets that did not become bare cores because of photoevaporation (the only formation path included in our model), but other mechanisms like a late formation after the dissipation of the disk such that they start with no H/He from the beginning \citep{Lee2014,LopezRice2016}, or a removal by impacts \citep[e.g.,][]{Schlichting2015}. Regarding the former scenario, the largest bare cores are indeed predicted to have radii of around 1.6 $\rearth$ \citep{LopezRice2016,Fulton2017}, consistent with the difference in the histogram of the rocky population and the actual planets. Also a more  efficient evaporation in some planetary systems than assumed here because of the spread in $L_{\rm XUV}$ and/or $\epsilon$ could lead to this group. We also note that the mean stellar radius in the \citet{Fulton2017} sample seems to be around 1.25 $R_\odot$, whereas we are only considering 1 $R_\odot$ stars. Another explanation is that these are  some bare icy cores. The position of the valley shows that icy cores cannot represent the dominant composition, but this does clearly not mean that there are no planets containing a lot of ice at all \citep[see also][]{Lopez2016}. They would tend to fill the valley preferentially near its lower boundary as visible from the right panel. Accurate density measurements of planets in the super-Earth peak distinguishing rocky from icy compositions will allow to break the degeneracy of the two possible explanations. For a Earth-like composition, the radii in the strongest dearth in the synthetic rocky population (1.3-1.8 $\rearth$) correspond to masses of about 3 to 9 $\mearth$ \citep[e.g.,][]{Mordasini2012b}. 

(4) A difference to the observed distribution that is common to both the rocky and icy core populations is an excess of large planets to the right of the sub-Neptune peak, i.e., at large radii. In the radius interval to the left of the sub-Neptune peak, evolutionary effects (evaporation) are of prime importance in sculpting the distribution. In contrast, to the right of the peak we more directly see the result of the formation, as evaporation is inefficient for these more massive planets. We also see that the core composition is less important for these larger planets, shown by the converging rocky and icy core distributions for radii larger than about 3-4 $\rearth$. This difference is therefore a direct consequence of the formation model that over-predicts intermediate size planets (and also hot Jupiters, see \citealt{Jin2014}). The reason could be a too efficient accretion of solids, too long synthetic disk lifetimes, etc. As it does not affect the main result (the location of the valley), this is however beyond the scope of this primarily evolutionary study.

(5) Another difference is the presence of an excess of planets in the super-Earth peak in the synthetic populations relative to the observations. This peak is quite high in the syntheses because it contains the cores of all planets that were evaporated out of the gap. In the rocky population, this excess becomes strong at radii of slightly more than 1 $\rearth$ (see also the radius distributions in \citealt{Jin2014}). Possible explanations are again an incorrect starting distribution predicted by the formation model (too many planets in the mass-distance interval that eventually become bare cores) or that effects other than atmospheric escape sculpt the distribution. 

We see that the ice mass fraction of planets that are in the distance-radius plane inside of triangle of evaporation is of high interest to disentangle the different explanations for the structure of the radius distribution, and to see whether the planets there indeed have a rocky composition as expected from the valley's position. This is addressed in the next section.

\section{Possible ice mass fractions of planets in the triangle of evaporation}\label{sect:icemassfraction}
In this section we use the results of the previous section on the location of the evaporation valley to derive constraints on the bulk composition of a sample of close-in low-mass planets.  We assume that planets that are located in the triangle of evaporation below the evaporation valley for rocky cores (the more conservative criterion) do not contain primordial H/He, i.e., that they are essentially solid planets consisting of iron, silicates in the form of  MgSiO$_{3}$ (perovskite), and potentially ices\footnote{These planets can potentially still have (secondary) atmospheres, but not thick primordial H/He envelopes that have the strongest impact on the radius because of the low molecular weight.}. The absence of H/He reduces the degeneracy in the mass-radius relation, an effect that was previously not included in a detailed way (using a coupled evolution and evaporation model) in similar analyses.

\subsection{{The iron mass fraction}}\label{subsect:ironmassfraction}
Unfortunately, even if H/He is absent and for vanishing observational errors, it is still not possible for a given mass and radius to derive in a unique way the ice mass fraction as the fraction of iron in the rocky part of the planet is in general still unknown. There are also differences introduced by silicates other than MgSiO$_{3}$, but these are in comparison of minor importance \citep[][hereafter SKHM07]{Seager2007}.  

However, in the Solar System, Earth, Venus, Mars, and Vesta all have a roughly chondritic bulk composition with relative mass fractions of silicates:iron of about 2:1 \citep[SKHM07,][]{AsphaugReufer2014}. Among the extrasolar planets, the mass-radius relation for planets with radii less than 2.7 $\rearth$ and with masses known with an error less than 20 \% are approximatively also compatible with such an Earth-like composition \citep{Dressing2015,Motalebi2015,Buchhave2016}.

An Earth-like iron mass fraction is also expected from condensation models for stars with a (scaled) solar composition, which is the typical chemical composition of stars in the solar neighborhood, at least for stars in the thin disk and [Fe/H] not too different from the solar value \citep{Santos2015}.

Furthermore, \citet[][hereafter GSS09]{Grasset2009} demonstrated that uncertainties related to Fe, Mg, and Si composition and temperature structure are secondary compared to the effect of the amount of water. As we are in this work interested only in the presence of large amounts of ices ($\sim$50\% in mass as expected for a formation beyond the ice line), and not a fine analysis of the composition, we assume that the rocky part of all planets has a 2:1 silicate:iron composition in mass.

It is clear that in the Solar System, Mercury with its massive metallic iron (about 70\% by mass), and the Moon with its small iron core, do not follow this relation. This shows that for individual planets, this assumption may not hold.  It is worthwhile mentioning that the objects in the Solar System with a clearly different composition are small bodies. Here, we only address planets with a radius of at least 0.75 $\rearth$.

\subsection{{The observational sample}}
For the analysis, we use the sample of \citet[][hereafter WM14]{WeissMarcy2014}  with 65 extrasolar planets smaller than 4 $\rearth$ with measured masses or mass upper limit both from radial velocity observations and TTVs. We exclude planets with a negative nominal mass and a radius of less than 0.75 $\rearth$ that have so large uncertainties in the mass that they cannot be used to meaningfully constrain the composition. We also exclude GJ 1214 b as our theoretical models apply to solar-like stars. With the exception of Kepler-138b which has an unconstrained density anyway, the other planets have host stars with masses between 0.75 and 1.25 $\msun$, clustering around 1 $\msun$. This leaves us with 55 planets.

Their semimajor axis $a$, the nominal mass $M$, the 1-$\sigma$ uncertainty in the mass $s_{\rm M}$, the radius $R$, and the 1-$\sigma$ uncertainty in the radius $s_{\rm R}$ are given in Table \ref{tablefice}. These values are directly taken from  WM14.

\begin{table*}[h]
\caption{Planetary characteristics, mean densities (in g/cm$^3$), inferred ice mass fraction and compositional type of planets in the NoDampf analysis of the WM14 sample outside (upper part) and inside of the triangle of evaporation (lower part). }
\begin{center}
\resizebox{\textwidth}{!}{%
\begin{tabular}{lcccccccc|ccc|cc}
 Name   &	$a$ [AU]	 & 			$M [\mearth]$   	&	$s_{M} [\mearth]$& $R [\rearth]$  &  $s_{R} [\rearth]$&	$\rho_{\rm min}$ &  $\rho_{\rm mean}$  & $\rho_{\rm max}$  &$f_{\rm ice,max}$&  		 $f_{\rm ice,mean}$&		 $f_{\rm ice,min}$ & $R/R_{\rm bare}$ & type \\ \hline
 Kepler-11c   &   0.107 &    2.90 &    2.20 &    2.87 &    0.06 &    0.15 &    0.68 &    1.27 &    1.00 &    1.00 &    1.00 &    1.83 &     1\\
 Kepler-11d   &   0.155 &    7.30 &    1.10 &    3.12 &    0.07 &    1.05 &    1.33 &    1.63 &    1.00 &    1.00 &    1.00 &    2.19 &     1\\
 Kepler-11f   &   0.250 &    2.00 &    0.80 &    2.49 &    0.06 &    0.40 &    0.71 &    1.08 &    1.00 &    1.00 &    1.00 &    1.99 &     1\\
 Kepler-30b   &   0.186 &   11.30 &    1.40 &    3.90 &    0.20 &    0.79 &    1.05 &    1.38 &    1.00 &    1.00 &    1.00 &    2.88 &     1\\
 Kepler-36c   &   0.128 &    8.10 &    0.53 &    3.68 &    0.05 &    0.80 &    0.90 &    1.00 &    1.00 &    1.00 &    1.00 &    2.46 &     1\\
 Kepler-79b   &   0.114 &   10.90 &    6.70 &    3.47 &    0.07 &    0.52 &    1.44 &    2.47 &    1.00 &    1.00 &    1.00 &    2.25 &     1\\
 Kepler-79c   &   0.184 &    5.90 &    2.10 &    3.72 &    0.08 &    0.38 &    0.63 &    0.91 &    1.00 &    1.00 &    1.00 &    2.74 &     1\\
 Kepler-79e   &   0.378 &    4.10 &    1.15 &    3.49 &    0.14 &    0.34 &    0.53 &    0.77 &    1.00 &    1.00 &    1.00 &    3.12 &     1\\
 Kepler-94b   &   0.034 &   10.84 &    1.40 &    3.51 &    0.15 &    1.06 &    1.38 &    1.78 &    1.00 &    1.00 &    1.00 &    1.64 &     1\\
 Kepler-95b   &   0.102 &   13.00 &    2.90 &    3.42 &    0.09 &    1.29 &    1.79 &    2.37 &    1.00 &    1.00 &    1.00 &    2.15 &     1\\
 Kepler-103b  &   0.128 &   14.11 &    4.70 &    3.37 &    0.09 &    1.25 &    2.03 &    2.94 &    1.00 &    1.00 &    1.00 &    2.25 &     1\\

 HD97658b     &   0.080 &    7.87 &    0.73 &    2.34 &    0.16 &    2.52 &    3.39 &    4.58 &    1.00 &    0.75 &    0.43 &    1.38 &     3\\
 Kepler-11b   &   0.091 &    1.90 &    1.20 &    1.80 &    0.04 &    0.62 &    1.80 &    3.14 &    1.00 &    1.00 &    0.62 &    1.10 &     3\\
 Kepler-20c   &   0.093 &   15.73 &    3.31 &    3.07 &    0.25 &    1.87 &    3.00 &    4.68 &    1.00 &    1.00 &    0.66 &    1.88 &     3\\
 Kepler-20d   &   0.345 &    7.53 &    7.22 &    2.75 &    0.23 &    0.06 &    2.00 &    5.08 &    1.00 &    1.00 &    0.47 &    2.40 &     3\\
 Kepler-25b   &   0.070 &    9.60 &    4.20 &    2.71 &    0.05 &    1.42 &    2.66 &    4.04 &    1.00 &    1.00 &    0.72 &    1.54 &     3\\
 Kepler-48c   &   0.085 &   14.61 &    2.30 &    2.71 &    0.14 &    2.93 &    4.05 &    5.49 &    1.00 &    0.74 &    0.43 &    1.62 &     3\\
 Kepler-68b   &   0.062 &    8.30 &    2.30 &    2.31 &    0.03 &    2.58 &    3.71 &    4.93 &    1.00 &    0.65 &    0.41 &    1.27 &     3\\
 Kepler-96b   &   0.126 &    8.46 &    3.40 &    2.67 &    0.22 &    1.16 &    2.45 &    4.45 &    1.00 &    1.00 &    0.55 &    1.77 &     3\\
 Kepler-100c  &   0.110 &    0.85 &    4.00 &    2.20 &    0.05 &   -1.53 &    0.44 &    2.69 &    1.00 &    1.00 &    0.91 &    1.41 &     3\\
 Kepler-102e  &   0.116 &    8.93 &    2.00 &    2.22 &    0.07 &    3.18 &    4.50 &    6.07 &    0.79 &    0.46 &    0.23 &    1.44 &     3\\
 Kepler-106c  &   0.111 &   10.44 &    3.20 &    2.50 &    0.32 &    1.78 &    3.69 &    7.26 &    1.00 &    0.73 &    0.08 &    1.61 &     3\\
 Kepler-106e  &   0.243 &   11.17 &    5.80 &    2.56 &    0.33 &    1.23 &    3.67 &    8.44 &    1.00 &    0.76 &    0.02 &    2.03 &     3\\
 Kepler-109b  &   0.069 &    1.30 &    5.40 &    2.37 &    0.07 &   -1.56 &    0.54 &    3.04 &    1.00 &    1.00 &    0.84 &    1.34 &     3\\
 Kepler-109c  &   0.152 &    2.22 &    7.80 &    2.52 &    0.07 &   -1.77 &    0.77 &    3.76 &    1.00 &    1.00 &    0.70 &    1.76 &     3\\

 Kepler-18b   &   0.045 &    6.90 &    3.48 &    2.00 &    0.10 &    2.04 &    4.76 &    8.35 &    1.00 &    0.34 &    0.00 &    1.01 &     4\\
 Kepler-48d   &   0.230 &    7.93 &    4.60 &    2.04 &    0.11 &    1.85 &    5.15 &    9.61 &    1.00 &    0.30 &    0.00 &    1.60 &     4\\
 Kepler-37d   &   0.212 &    1.87 &    9.08 &    1.94 &    0.06 &   -4.97 &    1.41 &    9.09 &    1.00 &    1.00 &    0.00 &    1.49 &     4\\
 Kepler-131b  &   0.126 &   16.13 &    3.50 &    2.41 &    0.20 &    3.92 &    6.36 &   10.03 &    0.72 &    0.28 &    0.00 &    1.60 &     4\\
 Kepler-409b  &   0.320 &    2.69 &    6.20 &    1.19 &    0.03 &  -10.66 &    8.80 &   31.41 &    1.00 &    0.00 &    0.00 &    1.02 &     4\\\hline

 CoRoT-7b     &   0.017 &    7.42 &    1.21 &    1.58 &    0.10 &    7.22 &   10.38 &   14.68 &    0.00 &    0.00 &    0.00 &    0.61 &     6\\
 Kepler-36b   &   0.115 &    4.46 &    0.30 &    1.48 &    0.03 &    6.66 &    7.59 &    8.61 &    0.00 &    0.00 &    0.00 &    0.96 &     6\\
 Kepler-68c   &   0.091 &    4.38 &    2.80 &    0.95 &    0.04 &    8.98 &   28.18 &   52.55 &    0.00 &    0.00 &    0.00 &    0.58 &     6\\
 Kepler-99b   &   0.050 &    6.15 &    1.30 &    1.48 &    0.08 &    7.05 &   10.46 &   14.97 &    0.00 &    0.00 &    0.00 &    0.77 &     6\\
 Kepler-100b  &   0.073 &    7.34 &    3.20 &    1.32 &    0.04 &    9.08 &   17.60 &   27.72 &    0.00 &    0.00 &    0.00 &    0.76 &     6\\
 Kepler-102d  &   0.086 &    3.80 &    1.80 &    1.18 &    0.04 &    6.07 &   12.76 &   20.85 &    0.00 &    0.00 &    0.00 &    0.71 &     6\\
 Kepler-131c  &   0.171 &    8.25 &    5.90 &    0.84 &    0.07 &   17.20 &   76.77 &  170.94 &    0.00 &    0.00 &    0.00 &    0.61 &     6\\

 55 Cnc e       &   0.015 &    8.38 &    0.39 &    1.99 &    0.08 &    4.94 &    5.86 &    6.99 &    0.34 &    0.20 &    0.04 &    0.75 &     7\\
 Kepler-48b   &   0.053 &    3.94 &    2.10 &    1.88 &    0.10 &    1.31 &    3.27 &    5.91 &    1.00 &    0.62 &    0.11 &    0.99 &     7\\
 Kepler-98b   &   0.026 &    3.55 &    1.60 &    1.99 &    0.22 &    1.00 &    2.48 &    5.12 &    1.00 &    0.94 &    0.23 &    0.86 &     7\\

 Kepler-10b   &   0.017 &    4.60 &    1.26 &    1.46 &    0.02 &    5.68 &    8.15 &   10.82 &    0.06 &    0.00 &    0.00 &    0.56 &     8\\
 Kepler-20b   &   0.045 &    8.47 &    2.12 &    1.91 &    0.16 &    3.95 &    6.70 &   10.90 &    0.51 &    0.06 &    0.00 &    0.96 &     8\\
 Kepler-37c   &   0.140 &    3.35 &    4.00 &    0.75 &    0.03 &   -7.55 &   43.80 &  108.61 &    1.00 &    0.00 &    0.00 &    0.51 &     8\\
 Kepler-78b   &   0.009 &    1.69 &    0.41 &    1.20 &    0.09 &    3.29 &    5.39 &    8.47 &    0.44 &    0.04 &    0.00 &    0.39 &     8\\
 Kepler-89b   &   0.051 &   10.50 &    4.60 &    1.71 &    0.16 &    4.98 &   11.58 &   22.36 &    0.28 &    0.00 &    0.00 &    0.89 &     8\\
 Kepler-93b   &   0.053 &    2.59 &    2.00 &    1.50 &    0.03 &    0.91 &    4.23 &    7.97 &    1.00 &    0.30 &    0.00 &    0.79 &     8\\
 Kepler-97b   &   0.036 &    3.51 &    1.90 &    1.48 &    0.13 &    2.13 &    5.97 &   12.13 &    0.98 &    0.03 &    0.00 &    0.70 &     8\\
 Kepler-102f  &   0.165 &    0.62 &    3.30 &    0.88 &    0.03 &  -19.61 &    5.02 &   35.20 &    1.00 &    0.02 &    0.00 &    0.63 &     8\\
 Kepler-106b  &   0.066 &    0.15 &    2.80 &    0.82 &    0.11 &  -18.17 &    1.50 &   45.46 &    1.00 &    1.00 &    0.00 &    0.46 &     8\\
 Kepler-113b  &   0.050 &    7.10 &    3.30 &    1.82 &    0.05 &    3.21 &    6.50 &   10.34 &    0.63 &    0.05 &    0.00 &    0.94 &     8\\
 Kepler-138b  &   0.012 &    0.06 &    1.20 &    1.07 &    0.02 &   -4.86 &    0.27 &    6.00 &    1.00 &    1.00 &    0.00 &    0.38 &     8\\
 Kepler-406b  &   0.036 &    4.71 &    1.70 &    1.43 &    0.03 &    5.33 &    8.88 &   12.88 &    0.10 &    0.00 &    0.00 &    0.68 &     8\\
 Kepler-406c  &   0.056 &    1.53 &    2.30 &    0.85 &    0.03 &   -6.23 &   13.74 &   38.31 &    1.00 &    0.00 &    0.00 &    0.45 &     8\\
 Kepler-407b  &   0.015 &    0.06 &    1.20 &    1.07 &    0.02 &   -4.86 &    0.27 &    6.00 &    1.00 &    1.00 &    0.00 &    0.40 &     8\\
 Kepler-408b  &   0.037 &    0.48 &    3.20 &    0.82 &    0.03 &  -24.43 &    4.80 &   41.17 &    1.00 &    0.04 &    0.00 &    0.39 &     8\\
 \end{tabular}
 }
\end{center}
{\footnotesize A $f_{\rm ice}$=1 means that H/He is necessary to explain a planet's density, not a 100\% ice composition. Types 1-4 are planets outside of the triangle of evaporation that should have kept H/He ($R/R_{\rm bare}$$>$1): (1) with H/He, (2) rocky (not occurring), (3) with H/He and/or ices, (4) unconstrained. Types 5-8 are planets in the triangle of evaporation  that should have lost all H/He ($R/R_{ \rm bare}$$<$1): (5) with H/He (not occurring), (6) rocky, (7) icy, (8) unconstrained.}
\label{tablefice}
\end{table*}%

\subsection{{Inferring the ice mass fraction}}
\begin{figure}
 \includegraphics[width=8.9cm]{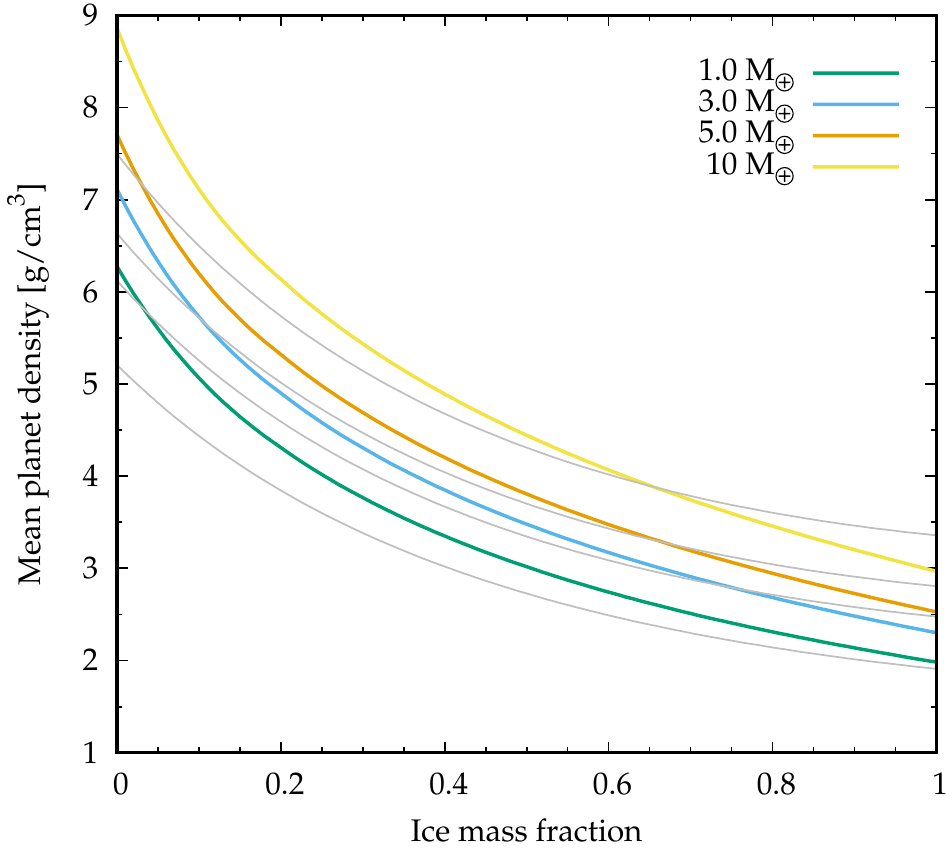}
 \caption{Mean density of solid planets with masses between 1 and 10 $\mearth$ as a function of the ice mass fraction. The rest of the planet has an Earth-like silicate and iron composition. Thick colored lines are obtained by integrating the modified polytropic EOS of \citet{Seager2007}, while the thin gray lines use the fits of  \citet{Grasset2009}. }
  \label{densfice}
\end{figure}

Using the masses and radii and their 1-$\sigma$ errors we then calculated the minimal density $\rho_{\rm min}=\rho(M-s_{\rm M},R+s_{\rm R})$, the mean density $\rho_{\rm mean}=\rho(M,R)$, and the maximal density $\rho_{\rm max}=\rho(M+s_{\rm M},R-s_{\rm R})$. These values are listed in Table \ref{tablefice}. We then used our internal structure model for solid planets \citep{Mordasini2012b} to derive the ice mass fraction that is needed to obtain these densities. This leads to the maximal ice mass fraction $f_{\rm ice,max}$, the mean $f_{\rm ice,mean}$, and the minimum $f_{\rm ice,min}$. Note that an $f_{\rm ice}=1$ in Table \ref{tablefice} means that H/He is necessary to explain a planet's density, and not a 100\% ice composition (no silicates and iron at all), as the planetary density is lower than the one obtained for a pure ice composition. As described in \citet{Mordasini2012b}, in this model we numerically integrate the equations of mass conservations and hydrostatic equilibrium using the modified polytropic EOS for iron, perovskite, and water ice of SKHM07 assuming a differentiated interior. For the rocky mass of the planet given as (1-$f_{\rm ice}) M $ we assume as mentioned a 2:1 silicate:iron composition by mass. The mean planetary density for planets with masses between 1 and 10 $\mearth$ as a function of $f_{\rm ice}$ is shown in Fig. \ref{densfice}. We first see the effect of self-compression with increasing planet mass. Second, we see that planets consisting of about 50\% ice, as expected for a formation outside of the iceline, have a density that is about half as high as for planets without ice. 

To quantify the sensitivity of the derived $f_{\rm ice}$ on the internal structure model, we have repeated the calculations using instead of the aforementioned modified SKHM07 polytropic EOS the fitting relations of GSS09 that yield the $f_{\rm ice}$ for given $M$ and $R$. These models are also shown in Fig.  \ref{densfice}. They use several more different EOS and explore the impact of various silicate compositions and temperature structures. They also find that even for masses and radii known without uncertainties, these factors allow to constrain the ice mass fraction to only about 5 \%. 

It is found that the derived ice mass fraction found for the planets in the WM14 sample with the two models agree relatively well, with differences in  $f_{\rm ice}$ of typically about 0.05 or less. For example, for 55 Cnc e, a mean $f_{\rm ice}$=0.204 is found with the SKHM07 polytropic EOS, while the GSS09 fits lead $f_{\rm ice}$=0.152. In  Table \ref{tablefice} we only show the results obtained with the SKHM07 model. In the statistical analysis below, we include both results. As we will see, using the two different models induces only minor changes in the major planet types identified, and does not change the statistical trends found. This is despite the  difference seen in Fig. \ref{densfice}. 

\subsection{Incorporating a vapor layer}
The two  internal structure models just described both assume that the water layer is in the solid form with material densities of about 1 g/cm$^3$ or higher. They thus neglect the radius enhancement resulting from the presence of a low-density vapor layer. To address this issue, we have considered two approaches to infer the water mass fraction:

In the NoDampf analysis, we neglect the vapor layer. Neglecting a possible low-density vapor layer means that the  ice mass fraction inferred in this analysis may be too high, as the vapor layer tends do reduce the mean density of the planet.

In the Dampf analysis, we take it into account in the following way: we first estimate the thickness of the vapor layer assuming it is isothermal with a temperature  that is equal  to the planet's equilibrium temperature $T_{\rm eq}$ for zero albedo, and extends from a low pressure $P_{\rm photo}$=20 mbar, the typical pressure level for the optical photosphere in a transit \citep{Lopez2016}, to  a high pressure $P_{\rm solid}$ where the density of  the vapor becomes  approximately unity, as assumed in the SKHM07 and GSS09 models. The equation of state ANEOS \citep{Thompson1990} shows that at the temperature of interest ($\sim10^3 K$ to order of magnitude) this should happen at pressures of $\sim10^{10}$ dyn/cm$^2$. The exact value is fortunately not important because of the weak logarithmic dependance.

The vapor layer thickness is then estimated as 
\begin{equation}
W_{\rm vap}=H  \ln\left(\frac{P_{\rm solid}}{P_{\rm photo}}\right)
\end{equation}
where $H=k_{\rm B} T_{\rm eq} / (\mu m_{\rm H} g)$ is the scale height with  $k_{\rm B}$ the Boltzmann  constant,  $\mu=18$ the vapor mean molecular weight, $m_{\rm H}$ the mass of  hydrogen, and $g$ the gravitational acceleration. 

Inserting these two pressures, one finds that the thickness is about 13 scale heights. We then subtract $W_{\rm vap}$ from the observed radius to obtain the radius $R_{\rm solid}$ of the solid part of the planet.  We  then  use $R_{\rm solid}$ and the (total) mass of the planet to again calculate the densities and  maximal,  mean, and minimal  $f_{\rm ice}$.  By subtracting  the vapor's  layer thickness, but neglecting its mass, we increase the planet's effective density, such that the ice mass fractions obtained in the Dampf analysis are lower than in the NoDampf  analysis. Note that the ice mass fractions found in this way are still not a strict lower limit, because of the isothermal approximation. But the difference between the Dampf and NoDampf analyses gives an measure how robust the results are.  

It is obvious that our analysis using a simple EOS or fits and only the  1-$\sigma$ errors does neither lead to accurate ice mass fractions for individual planets compared to more sophisticated EOS nor a full description of the consequences of the errors compared to, e.g., a Bayesian analyses \citep[e.g.,][]{Rogers2015,Dorn2017a}. But given the significant observational error bars that make it currently impossible to derive fine constraints even for well characterized exoplanets in any case \citep{Dorn2017b}, and our goal to reveal just the strongest statistical compositional tendencies and not the composition of individual planets, this approach is appropriate, as shown by the clear trends found below.   

\subsection{Planet classification}
For the classification, besides the three values of $f_{\rm ice}$, we also compare the planets' radius $R$ and semimajor axis $a$ with the local $R_{\rm bare}(a)=1.6 \times (a/0.1 \mathrm{AU})^{-0.27}$. If $R/R_{\rm bare}$ is larger than unity, we classify the planet as one outside of the triangle of evaporation (30 planets), and inside of it otherwise (25 planets). For planets outside of the triangle of evaporation, a H/He envelope is expected according to our theoretical evolution model. The quantity $R/R_{\rm bare}$ is given in the second last column of Table \ref{tablefice}.

The values of the three ice mass fractions and of $R/R_{\rm bare}$ finally allows us to classify the planets in the following 8 types. We always indicate the number of planets identified with the SKHM07 EOS for the NoDampf and, in parentheses, also the Dampf analysis. 

\begin{itemize}
\item Type I: outside, with H/He. These are planets with $f_{\rm ice,min}$ (and therefore also $f_{\rm ice,mean}$ and $f_{\rm ice,max}$) equal to 1 which means as mentioned  that H/He is needed to explain their low density, which is lower than for even for a pure ice composition. There are 11 (11) such planets in the sample.
\item Type 2: outside, rocky. These are planets where $f_{\rm ice,min}$ (and therefore also $f_{\rm ice,mean}$ and $f_{\rm ice,max}$)=0, i.e., which have a high density that does not allow the presence of ices or H/He. Interestingly, no such planets are present in both analyses, in agreement with the model predictions. This shows that planets outside of the triangle of evaporation have in all cases either kept H/He and/or contain ices.
\item Type 3: outside, with volatiles. These are planets with  $f_{\rm ice,min}<1$, but $>0$. This means that they have a density that is too low for a rocky composition. Volatiles are needed to explain them. As we are outside of the triangle of evaporation, it is not possible to constrain whether the volatiles are H/He or ices or a mixtures of both. The $f_{\rm ice}$ given for these planets are therefore upper limits. 14 (13) such planets are identified.
\item Type 4: outside, unclassified. These are planets with $f_{\rm ice,min}$=0 whereas $f_{\rm ice,mean}$ and/or $f_{\rm ice,max}$ are not zero, meaning that both rocky and volatile compositions are possible. This occurs when the density is too poorly constraint. 5 (6) planets.  
\item Type 5: inside, with H/He. This would be planets where H/He is necessary to explain their density ($f_{\rm ice,min}$=1). In agreement with the theoretical model which predicts that planets in the triangle of evaporation cannot keep their H/He, no such planets are found.
\item Type 6: inside, rocky. These are planets which have such high densities that $f_{\rm ice,min}$=0, i.e., which do not contain volatiles, but have a rocky composition. There are 7 (8)  such planets, one of which however has  nonphysically high densities.
\item Type 7: inside, icy.  These are planets  in the triangle of evaporation, that have a $f_{\rm ice,min}>0$, i.e., where ices are needed to explain their densities. There are three such planets in the NoDampf analysis (55 Cnc e, Kepler-48b, Kepler-98b) and none in the Dampf analysis. This  type of planet is particularly interesting for this work, and discussed further below.
\item Type 8: inside, unconstrained. These are planet with an unconstrained composition, as they have $f_{\rm ice,min}=0$ and $f_{\rm ice,max}>0$. With 15 (17) planet, this group is the the largest, illustrating the difficulty to observationally obtain masses of such small planets that are sufficiently precise to constraint the composition.
\end{itemize}

\begin{figure*}
\begin{center} %[width=15cm]
 \includegraphics[width=0.9\textwidth]{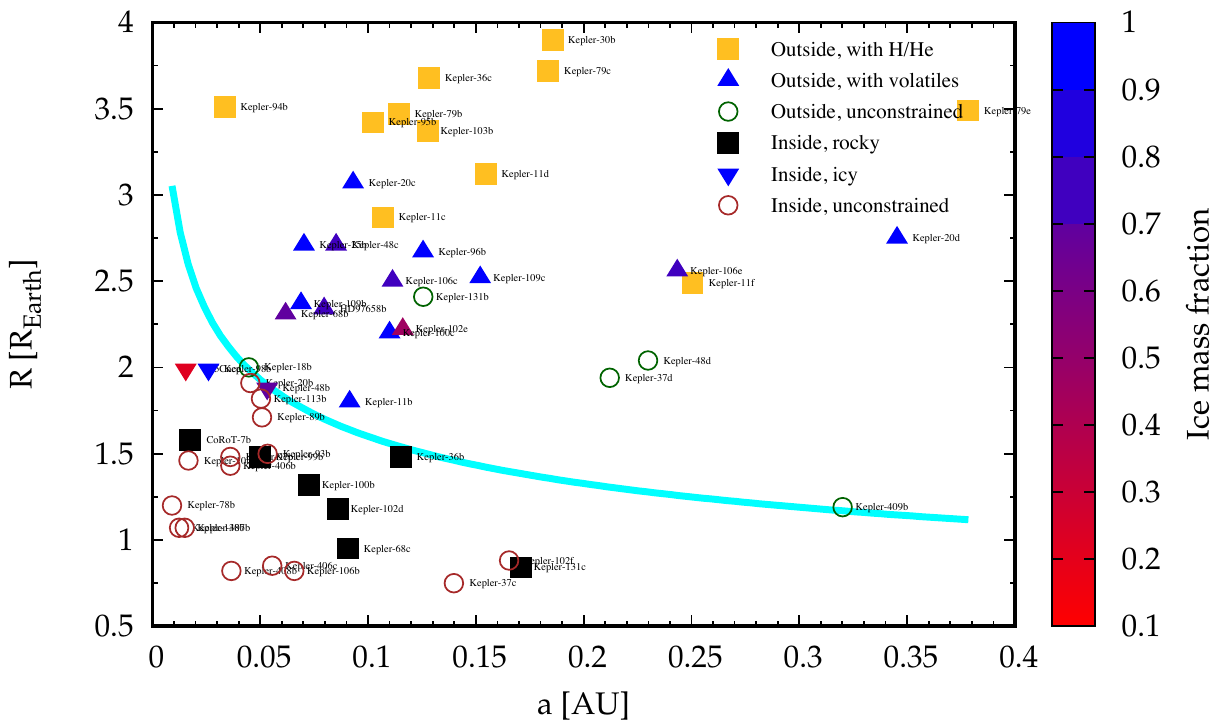}
 \caption{Planetary type and ice mass fraction $f_{\rm ice,mean}$ (color coded) as a function of distance and radius for planets in the NoDampf analysis of the WM14 sample.  In this analysis, the thickness of a possible vapor layer is neglected, leading to higher inferred ice mass fractions. Planets below the cyan line are in the triangle of evaporation. Under the assumption that rocky material has a 2:1 silicate-to-iron mass ratio, one finds six planet type based on the position relative to the cyan line and the mean density. Outside of the triangle, three types are identified: Yellow squares: Type 1, planets with H/He. Color coded upward triangles: Type 3, planets with  H/He and/or ices. For these planets, the indicated ice mass fraction is an upper limit as they can also contain H/He. Open green circles: Type 4, unconstrained composition because of too large uncertainties in the density. Inside of the triangle of evaporation: black squares: Type 6, rocky composition. Downward pointing triangles: Type 7, icy composition. Open brown circle: unconstrained composition. The types 2 (outside, rocky) and 5 (inside, with H/He) do not occur, in agreement with the theoretical model. The figure indicates a predominantly rocky composition in the triangle of evaporation, and thus a formation inside of the iceline.}
  \label{RcompoNoAtmoSubtr}
\end{center}
\end{figure*}

\begin{figure*}
\begin{center} %[width=15cm]
 \includegraphics[width=15cm]{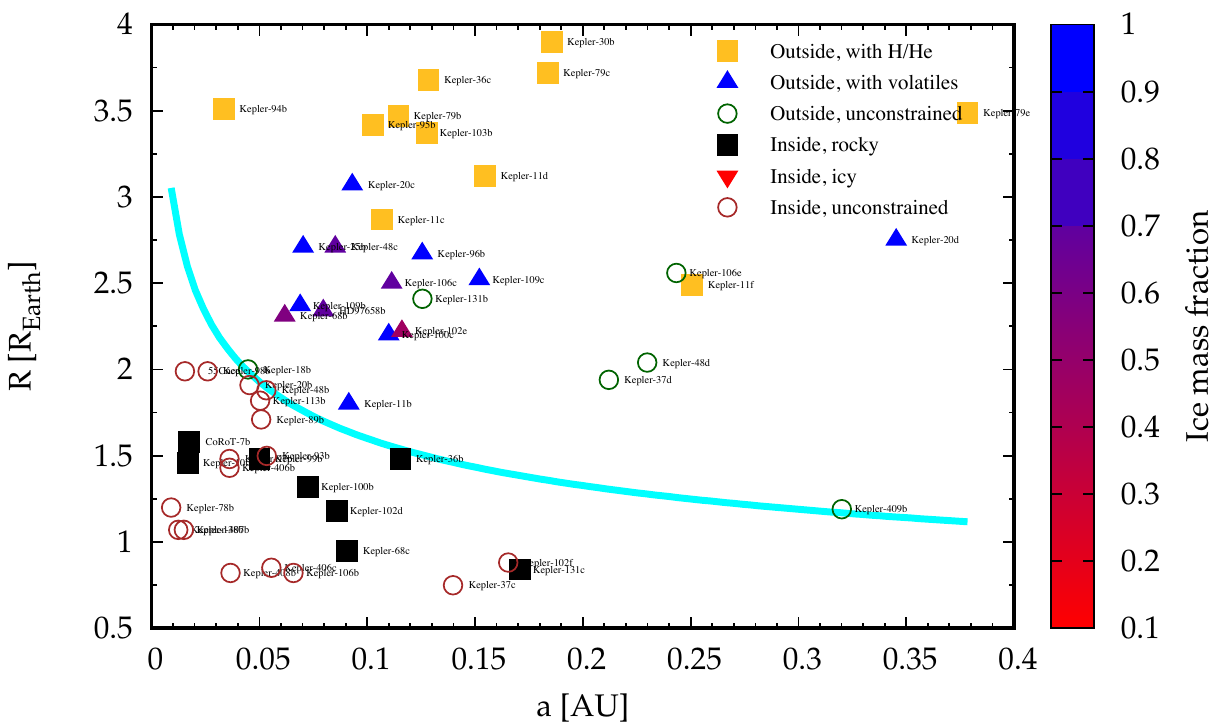}
 \caption{Analogous to Fig. \ref{RcompoNoAtmoSubtr}, but in the Dampf analysis, i.e.,  taking into account the thickness of a possible isothermal vapor layer. This yields lower estimates of the ice mass fraction.  The general trend is the same as in Fig. \ref{RcompoNoAtmoSubtr}. But the three planets (55 Cnc e, Kepler-48b, Kepler-98b) that were classified in the NoDampf analysis as Type 7 (inside, icy) are now unconstrained. This shows that there is currently no secure detection of a planet in the triangle of evaporation with a water-dominated composition.}
  \label{RcompoAtmoSubtr}
\end{center}
\end{figure*}

The result of the NoDampf analysis with the SKHM07 EOS of the WM14 sample regarding these 8 types is given in the last column of Table \ref{tablefice} and visualized in Fig. \ref{RcompoNoAtmoSubtr}. Figure \ref{RcompoAtmoSubtr} shows the results also with the SKHM07 EOS, but in the Dampf analysis, i.e., with the effect of the vapor layer. 

%\centering
%\begin{minipage}{0.6\textwidth}
%\end{minipage}

Both figures show a clear compositional gradient with increasing planet radius which is in general agreement with earlier studies \citep[e.g.,][]{Marcy2014,Rogers2015,WolfgangLopez2015}: for radii less than about 1.6 $\rearth$, we find rocky compositions. At radii between about 1.6 and 3 $\rearth$, volatiles are required, but it is not constrained whether it is H/He and/or ices. Finally, for  $R\gtrsim 3 \rearth$, H/He is usually required to explain the density. The theoretically predicted transition to rocky planets given by the cyan line is in both analyses broadly speaking consistent with the location in the observational data, but in this small sample here it is difficult to derive more precise constraints. For this, the larger \citet{Fulton2017} sample is more constraining. In contrast to the clear dependency on the radius, from the distribution of the observed planet types in Fig. \ref{RcompoNoAtmoSubtr} and \ref{RcompoAtmoSubtr} it is not obvious that there is also a gradient to more volatile compositions with orbital distance, as predicted by the evaporation model. Determining this observationally would be very important for example with CHEOPS \citep{Broeg2013} and later PLATO 2.0 \citep{Rauer2013}.

A positive agreement between theory and observation is that neither Type 2 (outside, rocky) nor  Type 5 (inside, with H/He) are identified, which would be in contraction to the theoretical model that explains the distance-radius structure by a scenario where at post-formation time all planets have H/He, with the planet in the triangle of evaporation losing the H/He in the subsequent evolution making their bare rocky cores visible, and the others keeping it. Note that it could still be possible that "above" of the triangle of evaporation at higher masses and orbital distance where we cannot probe the core composition, planets have an icy core below the H/He envelope. Such a compositional gradient is  predicted by the formation models of \citet{Alibert2013}.

Our result of a clear compositional trend was obtained under the simplifying assumption of an Earth-like silicate:iron fraction in all planets. This is not expected if the simplification would dominate the results. The underlying reason why the assumption does not blur the trend are the strong density changes induced by adding large amounts of ices and even more so by H/He compared to the modest changes introduced by varying the  silicate:iron fraction over a plausible range.

\subsection{{Individual planets}}
The most important question we wanted to address in this section is whether there are clearly ice-dominated planets in the triangle of evaporation i.e., below the cyan line (Type 7 planets). A dominance of such planets would be in contradiction to the results of Sect. \ref{rockyicy} where it was found that the location of the evaporation valley is consistent with mainly rocky, but not icy cores.  In the NoDampf analysis, three planets are found to be of Type 7, but none in the Dampf analysis.

The first is 55 Cnc e \citep{McArthur2004,Demory2011}. For it, a $f_{\rm ice,mean}$=20\% with an interval between 4 to 34\% ice was found, in agreement with \citet{Demory2011}. However, it is well known that significantly different planetary properties have been reported observationally for this planet with strong consequences for the inferred composition (see discussion in \citealt{Dorn2017b}). Furthermore, in the Dampf analysis, we find that the maximal, mean, and minimal  $f_{\rm ice}$ are 22, 5, ad 0\%, showing that in the Dampf analysis, 55 Cnc e's composition is unconstrained. Thus, for 55 Cnc e an ice-dominated composition cannot be firmly established. 

The second planet is Kepler-48 b \citep{Marcy2014}. For this object we note that $R/R_{\rm bare}(a)$=0.99, i.e., it is only just inside of the triangle of evaporation, and $(R+s_{\rm R})/R_{\rm bare}(a)$ is even bigger than unit (1.04). More importantly, in the Dampf analysis, its composition is again unconstrained (1.0, 0.46, 0 for the the maximal, mean, and minimal $f_{\rm ice}$, respectively). Again, a clearly ice-dominated composition cannot be established.

Finally, there is Kepler-98 b. Its $R/R_{\rm bare}(a)$ is 0.86, i.e. it is further away from the boundary than Kepler-48 b, but its large $s_{\rm R}$ (the largest of all planets in the triangle) make that $(R+s_{\rm R})/R_{\rm bare}(a)$ is 0.96.  As for the previous two cases, the Dampf analysis gives in contrast an unconstrained composition, with a maximum, mean, and  minimal $f_{\rm ice}$=1.0, 0.58, 0.0.

All these results are identical for both the SKHM07 and GSS09 EOS. In summary this means that we could not identify a secure water-dominated composition for any planet in the triangle of evaporation.

\subsection{Statistical analysis}
In Tables \ref{statsnodampf} and \ref{statsdampf} we report the number and percentage of the different planet types in the NoDampf and Dampf analyses. Results for both the SKHM07 and GSS09 models are given such that we can compare the results of four different classification methods, allowing to see how sensitive the results are to models assumptions. 

\begin{table}[h]
\caption{Number and percentage of planet types using the EOS of 
SKHM07 (left) and GSS09 (right) without a vapor layer (NoDampf analysis)}
\begin{center}
\begin{tabular}{lcc|cc}
Quantity & Nb.  & \% & Nb. &  \%\\ \hline
Outside of  triangle       &       \multicolumn{4}{c}{30}  \\
Type 1 (with H/He)  &            11  & 37 &  12 & 40\\   
Type 2 (rocky)     &       0  &  0  & 0 & 0\\  
Type 3 (with volatiles)    &      14 & 46 & 12 & 40 \\    
Type 4 (unconstrained)        &       5 & 17  & 6 & 20 \\  \hline  
Inside of triangle       &       \multicolumn{4}{c}{25}  \\
Type 5 (with H/He)            &   0  & 0  & 0 & 0\\    
Type 6 (rocky)          &   7 & 28  & 7  & 28\\    
Type 7 (icy)           &    3 & 12 & 3 & 12 \\    
Type 8 (unconstrained)           &    15 & 60 & 15 & 60  \\
 \end{tabular}
 \end{center}
\label{statsnodampf}
\end{table}%

\begin{table}[h]
\caption{Number and percentage of planet types using the eos of SKHM07  (left) and GSS09 (right) with a vapor layer (Dampf analysis)}
\begin{center}
\begin{tabular}{lcc|cc}
Quantity & Nb.  & \% & Nb. &  \%\\ \hline
Outside of  triangle       &       \multicolumn{4}{c}{30}  \\
Type 1 (with H/He)  &            11  & 37 &  11 & 37\\   
Type 2 (rocky)     &       0  &  0  & 0 & 0\\  
Type 3 (with volatiles)    &      13 & 43 & 13 & 43 \\    
Type 4 (unconstrained)        &       6 & 20  & 6 & 20 \\  \hline  
Inside of triangle       &       \multicolumn{4}{c}{25}  \\
Type 5 (with H/He)            &   0  & 0  & 0 & 0\\    
Type 6 (rocky)          &   8 & 32  & 8  & 32\\    
Type 7 (icy)           &    0 & 0 & 0 & 0 \\    
Type 8 (unconstrained)           &    17 & 68 & 17 & 68  \\
 \end{tabular}
\end{center}
\label{statsdampf}
\end{table}%

One first notes that the statistical results using the SKHM07 and GSS09 internal structure models only vary little for the NoDampf analysis, and that they are even identical for the Dampf analysis. Given the differences in Fig. \ref{densfice}, this may appear somewhat surprising at first, but it reflect that these differences are of second importance relative to the observational errors and the large density change induced by a H/He layer and the presence of a large amount of ice relative to no ice at all. Second, we see that expect for Type 7 (inside, icy) also the Dampf and NoDampf analyses give very similar results. This is positive as it again indicates that the statistical results are not strongly affected by specific model setting.

Coming back to the question about the composition of planets in the triangle of evaporation, from the number of planets of Type 6 (inside, rocky) and Type 7 (inside, icy) in Tables \ref{statsnodampf} and \ref{statsdampf}, we see that taken at face value, for the planets with a constrained composition, between 70-100\% of the planets in the triangle of evaporation have a rocky composition, and 0-30\% have an icy composition, with a value likely closer to zero. It is clear that these values are derived from a small sample, with a fixed iron:silicate ratio, and without a analysis of the errors that goes beyond the 1-$\sigma$ uncertainties. But they nevertheless hint at a  predominantly rocky composition of planets in the triangle of evaporation.  

In summary we have found two main results in this section: first, that there is a clear trend from a rocky composition at radii less than about 1.6 $\rearth$ over a volatile-rich composition with ices and/or H/He at intermediate radii (1.6-3 $\rearth$) to one with H/He for even larger radii. The dependency on orbital distance in the observations is in contrast unclear. Second, that we could not find individual planets with a robustly volatile-dominated composition in the triangle of evaporation, but that the planets there with a sufficiently well known density have a rocky composition. This agrees with the statistical result of the location of the evaporation valley in Section \ref{rockyicy} that also points towards a rocky composition in the bare core triangle. 

\section{The Planetary Mass--Mean Density Diagram as a Function of Distance and Time}\label{massdensity}
  
\begin{figure*}[htb]
\begin{center}
 \includegraphics[width=13.5cm]{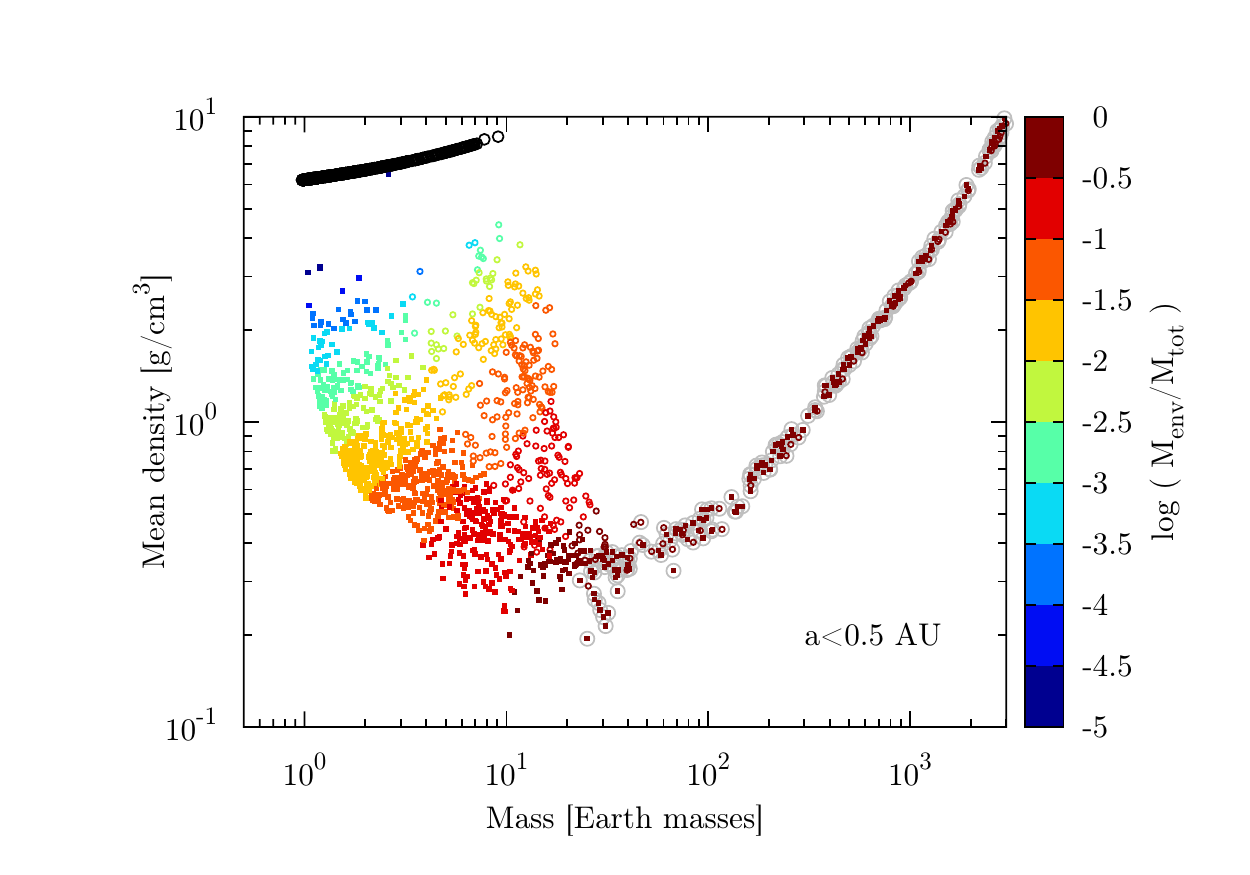}
 \caption{Mass versus mean density for the synthetic population with rocky cores at an age of 5 Gyr {exhibiting the characteristic  broken V-shape}. The colors indicate the mass fraction of the H/He envelope at this time. Larger gray circles additionally show when $M_{\rm env}/M_{\rm tot}>0.5$, i.e., it shows the transition to gas-dominated planets. Black symbols are planets that have lost all primordial H/He. The shape of the colored points represents a planet's semimajor axis: open circles if $0.06<a$/AU$<0.15$ and filled squares if $0.15<a$/AU$<0.5$.   }
  \label{MDoverview}
\end{center}
\end{figure*}  
  
The planetary mass--mean density diagram highlights the structural difference shown in the $a$--$R$ distribution \citep{Rauer2013,HatzesRauer2015}. Compared to the $M$--$R$ diagram, it shows compositional changes more clearly because of its 1$/R^{3}$ dependency. The radius $R$ itself is in comparison only a weak function, changing only by a factor $\sim20$ for planetary masses varying over four order of magnitude. If accurate measurement of planetary ages are available, one of the major scientific goals of the PLATO 2.0 satellite \citep{Rauer2013}, planetary mean density in time can be a novel component to reduce the degeneracy in different planetary bulk compositions.

\subsection{General structure of the synthetic mass-density diagram}
As an illustration of the general structure of the synthetic mass-density diagram, Figure \ref{MDoverview} shows the  $M$--$\bar{\rho}$ diagram of the  synthetic planetary population with rocky cores for $0.06<a$/AU$<0.5$ at an age of 5 Gyr. 

Its {characteristic broken V-shape \citep{Rauer2013} reveals} several structures that are related to both planet formation and evolution. In order of increasing mass, they are:
\begin{enumerate}
\item The black points in the top right corner are low-mass planets that have either started without a H/He envelope or have lost it due to evaporation (as it is the case in the model). In this plot, these solid planets follow a single mass-density relation because a universal 2:1 silicate:iron ratio was adopted for this population. 

\item An empty evaporation valley separate these solid planets from planets that retain an H/He envelope{, breaking the ``V'' into two parts}. It is the same valley as found in the $a$--$R$ plot, but it is more clear in the mean-density plot since a  small amount of H/He (just 0.1\% in mass) can already decrease the mean-density by a factor of 2-3. Given the result of \citet{Fulton2017} it is expected that once we have a sufficiently high number of planets in this region with well constrained $\bar{\rho}$ such a underpopulated valley should also appear in the observational mass--density diagram.
 
\item Another empty region that is also a result of evaporation is the bottom left corner. This region remains empty because only planets inside 0.5 AU are included in this plot, and we are at a late moment in time, 5 Gyr. At such close-in orbits, low-mass planets with very low mean densities  quickly lose all their gaseous envelopes due to intense evaporation on a short timescale. Hence they have moved to higher mean densities.

\item The planets that retain a H/He envelope form a {(continuous)} V-shape that is related with the core-accretion model \citep{Rauer2013,Baruteau2016}. In the left branch of the ``V'', i.e. for low-mass core-dominated planets with H/He corresponding to (sub-)Neptunian planets, the most distinct feature is that their location in the mass-density plot reveals their envelope mass-fractions indicated in the figure by the color \citep{Lopez2014,Jin2014}. The left part of the V-shape also shows the effect of evaporation for close-in low-mass planets: at a given total mass, the hotter a planet (the closer it is to the star as indicated by the symbols' shape), the higher its density since more primordial H/He was lost due to stronger evaporation. This indicates that mainly evolution (evaporation) and not formation shapes this region at least for the small orbital distances we consider here \citep{Owen2013}. We discuss this in more details below.

\item The most notable feature in the plot is a change of regime at about 0.1 $M_{\rm Jupiter}$ ($\sim$ 30 $M_{\oplus}$). At this mass, for planets with (remaining) primordial H/He, the density changes from decreasing with increasing mass because of an increasing H/He mass fraction, to a typical density that is first only weakly dependent on mass (for $M\lesssim 70 M_\oplus$), to finally a density that  increases with mass, because of the increasing self-compression of the gas. The lowest $\bar{\rho}$ {occur for} planets {with} 10-30 $M_{\oplus}$. {Particularly low $\bar{\rho}$ correspond to planets with the highest envelope mass fraction that are in the outer part of the considered orbital distance interval. The formation track of such planets was such that the core accretion rate and thus luminosity was low towards the end of the disk lifetime, making  a more efficient gas accretion possible \citep[e.g.,][]{Ikoma2000}. This} mass range {also} corresponds to the transition point where rapid gas accretion starts in the core-accretion scenario \citep{Pollack1996}, and planets with masses beyond this range will become gas-dominated. This is indicated by the gray circles in the figure.  The right part of the V-shape thus shows the gas-dominated giants.

\item At the highest masses ($\gtrsim100-200\mearth$), the density finally increases linearly with mass, as expected for a n=1 polytrope that provides a reasonable approximation to the internal structure of giant planets in this mass domain \citep{Baruteau2016}. Note that the synthetic mass--density relation of gas giants in the figures is artificially sharpened in the synthetic populations for two reasons. First, we do not include bloating mechanisms like, e.g., ohmic heating \citep{Batygin2011}. Second, in our model all planets use the same opacity laws (solar composition opacity during evolution), while in reality the compositions of the planetary atmospheres and thus the opacity will vary \citep[e.g.][]{Mordasini2016,Espinoza2016}. This in turn affects the cooling and hence the planetary radius \citep[e.g.,][]{Burrows2011,Vazan2013}. The lack of bloating mechanisms explains why the minimal density in the synthetic populations is around 0.4 g/cm$^3$ for giant planets, while in the actual population, there are giant planets with a mean density that is about a factor three lower (see Fig. \ref{MDcomp}). Interestingly, in the observational data, there are also giant planets that have, at a given mass, a mean density that is clearly higher than in the synthetic population, which is caused by heavy element contents higher than in the synthetic counterparts \citep[e.g.,][]{Leconte2011}. As can be seen from Fig. \ref{MDevo} below, at larger orbital distances  ($\gtrsim5$ AU) there are are synthetic planets with such higher densities. This is an indication that the theoretical model does not predict close-in giant planets with sufficiently high enrichments. This could be a compositional indication that effects other than disk migration which is the only process considered in the formation model (also) lead to giant planets \citep[e.g.,][]{Crida2014}.

\end{enumerate}

\subsection{The mass-density diagram in time}
\begin{figure*}[htb]
 \includegraphics[width=17.5cm]{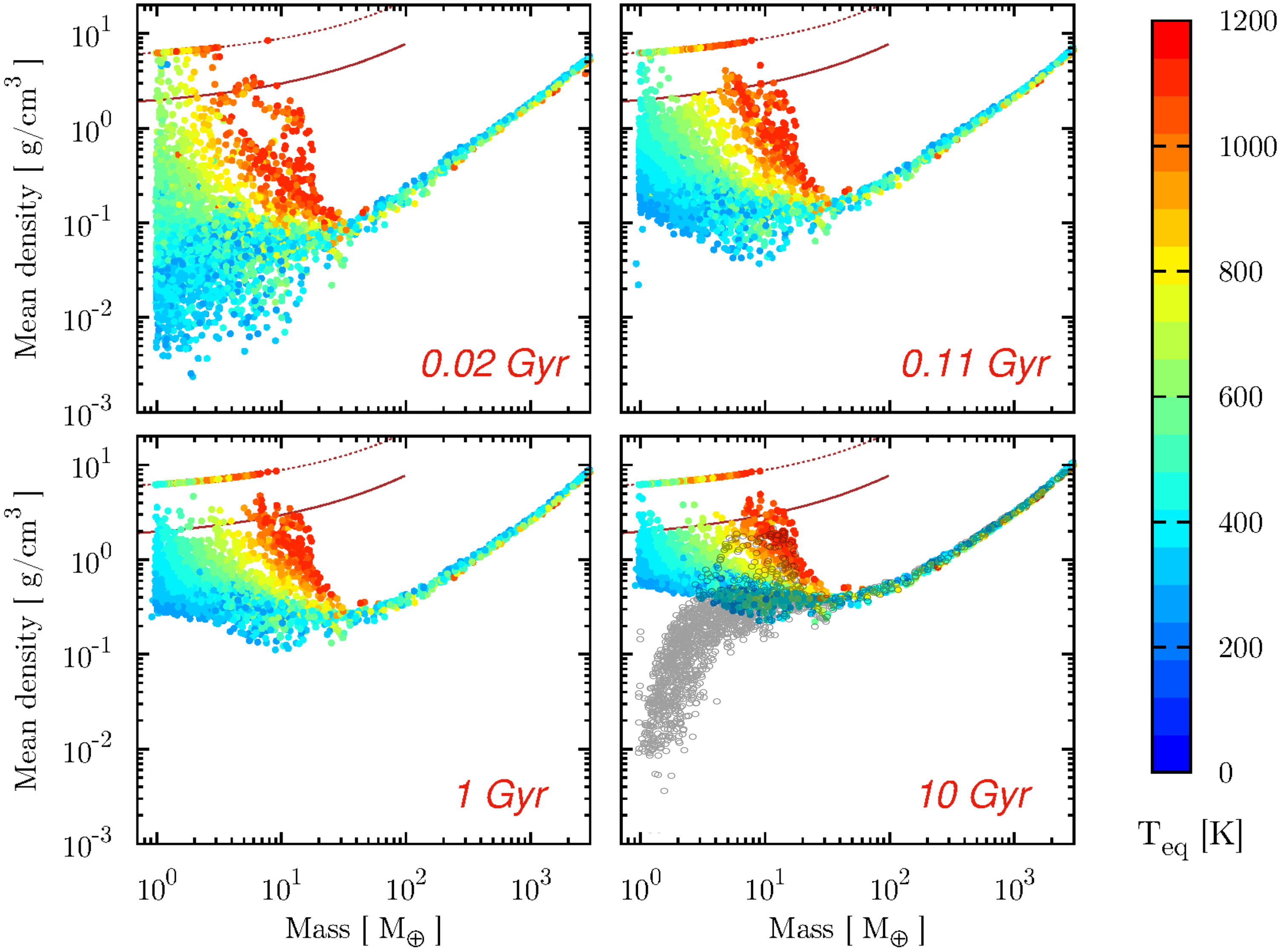}
 \caption{Temporal evolution of the planetary mass vs. mean density of  planets between 0.06--1 AU in the  population with rocky cores. The color of each point shows the equilibrium temperature of the planet. The dotted and solid lines in the top left corner show the densities of Earth-like and icy cores without H/He, respectively. Note that colors of the planets on the dotted line at for example 1 Gyr give the incorrect impression that there are no, e.g., 1 $\mearth$ planets hotter than about 400 K on that line. In reality, the hotter planet are hidden ``under'' the colder ones by the plotting method. For planets not on the line, there is in contrast no such misleading covering-over, but the colors indicate the real correlation that at fixed total mass $\lesssim 30  \mearth$, hotter planets  have a higher density. In the panel at 10 Gyr, gray open symbols show the same population neglecting atmospheric escape. Note the general contraction of the planets as well as how atmospheric escapes eliminates in time warm and hot low-density planets of low mass. }
  \label{MDtemp}
\end{figure*}

As mentioned in the introduction, the evolution of planetary radii in time could be a way to constrain their composition, to break or reduce the degeneracies, and to thus better understand their nature (gaseous, solid, icy, rocky). Figure \ref{MDtemp} shows the temporal evolution of the mass--density distribution of the planets between 0.06--1 AU in the rocky core population. The color gives the planetary equilibrium temperature. 

One notes how the mean densities of the planets with H/He increases in time. This is for the giant planets in the right part of the ``V'' mostly due to cooling and contraction at constant mass. It causes the densities of these gas-dominated planets to increase by about a factor 2-3 from 20 Myr to 10 Gyr.  Most of the contraction happens early-on, but significant changes still occur between 1 and 10 Gyr, the observationally more accessible time interval. For close-in core-dominated planets with H/He in the left part of the ``V'', evaporation is the dominant effect shaping the density in the interval of orbital distance that we consider. It can lead to an increase of the mean density by up to a factor $\sim$100.

Planets without H/He in contrast do not undergo significant changes of their mean density in time. This means that determining observationally whether the mean density of a certain sub-group of planets (in a interval in mass and insolation) changes between ages of 0.1 and 5 Gyr allows to see whether they contain H/He. 

The features produced by evaporation are clearly visible in the left part of the panels in Figure \ref{MDtemp}. Shortly after the end of formation, at 0.02 Gyr, some low-mass planets in the left bottom corner have very low densities $\lesssim10^{-2}$ g/cm$^{2}$ as a tenuous envelope can produce a large increase in the planetary radius \citep[e.g.,][]{Adams2008,Jin2014}. Because the extended envelopes of such hot low-density planets are rapidly removed by evaporation (the evaporation rate in the energy limited domain scales as $1/\bar{\rho}$), the low-mass very low-density planets at 0.02 Gyr disappear in the snapshots at later times.

This produces a large number of close-in low-mass planets that have been evaporated to bare rocky cores. These bare rocky cores lie on the Earth-like mass--radius relationship (dotted curve) as all of them have an identical 2:1 silicate:iron ratio. One sees how the most massive planet on the dotted line increases in time, as more massive (and colder) planet lose their envelope later. 
 
Most of the density changes because of evaporation happen between the panel at 20 Myr and 110 Myr when the stellar $L_{\rm XUV}$ is high and the planetary radii are large. The snapshots at 1 and 10 Gyr show almost the same gap in the mass--density space, the only difference between them is that the densities of the planets that still retain an envelope increase at the 10 Gyr's snapshot due to planet cooling. The fact that most temporal changes in the mass--density diagram happen in the first 0.1 Gyr mean that it is more difficult to directly observe them as most (bright) stars are older than this. Some temporal change however still happens between the panels at 1 and 10 Gyr, but it requires more precise measurements.

In the panel at 10 Gyr, we also show the same population but neglecting evaporation. This means that even very low-mass hot planets (artificially) keep all their primordial H/He. The weak gravity, the strong stellar irradiation and associated high planetary temperatures and large scale heights mean that these planets  have very large radii and extremely low densities. An increase of the radius with decreasing mass for hot low-mass planets is a well known effect \citep[e.g.,][]{Rogers2010density,Mordasini2012b}. In reality, evaporation extremely would quickly removes such envelopes.

Comparison with observations (Fig.   \ref{MDcomp}) shows that such a scenario without evaporation can be ruled out. This illustrates again that for such planets, evaporation plays a decisive role in shaping their radii \citep{Owen2013}. 

\begin{figure}
\begin{center}
 \includegraphics[width=0.48\textwidth]{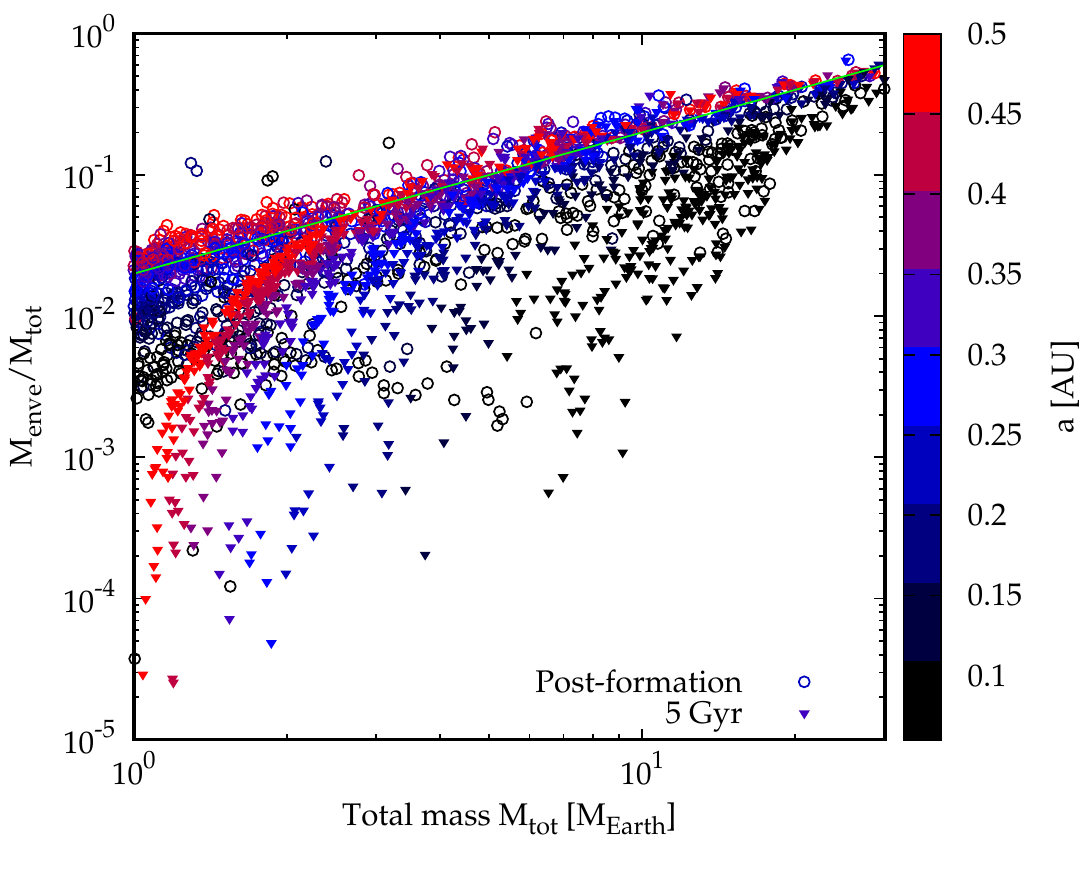}
 \caption{H/He envelope mass fraction as a function of total mass for low-mass synthetic planets with 0.06$<a/$AU$<$0.5, immediately after formation (open circles) and at 5 Gyr (filled triangles), color-coded according to orbital distance. The green line shows At the lowest masses, evaporation completely removes the envelopes. At higher masses, it induces a stronger spread, depending on distance. It leads in particular to planets of $\sim$5-10$\mearth$ with little H/He, much less than typically after formation.}
  \label{MenvefractionAW}
  \end{center}
\end{figure}

\subsection{{Evolution of the envelope mass fraction}}
Figure  \ref{MenvefractionAW} shows the H/He envelope mass fraction in the synthetic population as a function of total mass for low-mass planets, again for 0.06$<$$a/$AU$<$0.5. The population is shown immediately after the end of formation (open circles) and at 5 Gyr (filled triangles). For both sets, the colors show the semimajor axes. The primordial envelope mass fraction as a function of total mass  is an important constraint for formation  models. It depends on the opacity in the protoplanetary atmosphere \citep{Podolak2003,Ormel2014,Mordasini2014}, the orbital distance during formation \citep{Ikoma2012}, and the planet's accretional heating \citep{Ikoma2000}. The green line plots  $M_{\rm enve}/M_{\rm tot}=0.02 M_{\rm tot}/\mearth$ to guide the eye. For this population, an atmospheric grain opacity during formation reduced by a factor 0.003 relative to ISM opacities was assumed \citep{Mordasini2014}. The line indicates how the primordial envelope mass typically increases with total mass because of the shorter KH-timescales of more massive planets. The linear increase means that the these planets  have an effective KH-timescale that scales as 1/$M_{\rm tot}$ \citep{Mordasini2014}. The colors show that already during formation, there is also a positive correlation of envelope mass and orbital distance \citep{Ikoma2012,Lee2015}. A dependency roughly $\propto a^{0.7}$ is found in the synthesis, but with a lot of scatter, originating mostly from the solid accretion rate and thus luminosity at the time of gas disk dispersal. 

\begin{figure*}
 \includegraphics[width=17.5cm]{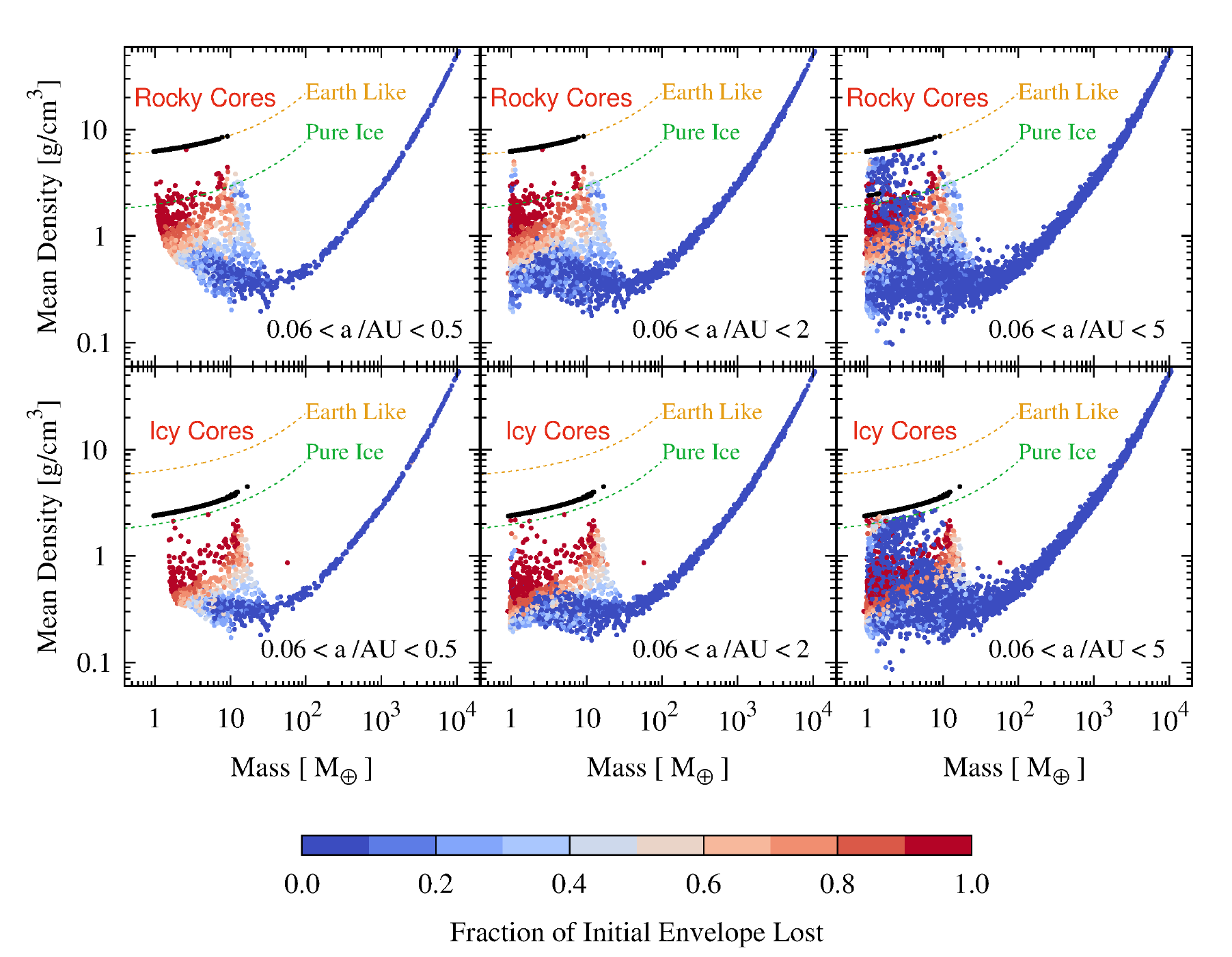}
 \caption{Planetary mass vs. mean density of the rocky (top row) and icy core (bottom row) populations at 5 Gyr for different orbital distances. The color of each point shows the fraction of the initial envelope that was evaporated. The black points are planets that have lost all their initial envelope. The orange and green dashed lines show the density curves of Earth-like and pure-ice cores respectively (``icy'' in contrast means 75\% ice). For rocky and icy cores there is a dearth of planets with densities of 3-7 and 1-3  g/cm$^{3}$, respectively, reflecting the different loci of the evaporation valley. With increasing distance, low-mass very low-density planets as well as planets in the evaporation valley appear.  }
  \label{MDevo}
\end{figure*}

We see how evolution in the form of evaporation modifies the primordial $M_{\rm enve}/M_{\rm tot}$ in two ways:  first, at the lowest masses, the envelopes are completely removed or so strongly reduced that they are always much smaller than directly after formation. For higher masses, evaporation induces a stronger spread in the envelope mass fraction at a given total mass compared to formation. Depending on distance, some planets still have envelope masses comparable to the primordial mass, but there are also planets where only 1\% (or less) of the primordial mass is left. This in particular means that there are some relatively massive planets (5-10 $\mearth$) with only very little H/He ($M_{\rm enve}/M_{\rm tot}\sim10^{-3}$). After formation, such planets rather have $M_{\rm enve}/M_{\rm tot}\sim10^{-1}$.  These points explains the differences in the mean densities between the populations with and without evaporation shown in the 10 Gyr panel of Fig. \ref{MDtemp}. 

Gas accretion during formation, and gas loss during evolution follow the same trend (less envelope for lower-mass, closer planets), although the scalings are different. This makes it more difficult to disentangle the two effects. At younger ages, at larger orbital distances, and for more massive planets, the imprint of formation and in particular the way how the KH timescale depends on mass is therefore more clearly preserved.

\subsection{Impact of the core composition and orbital distance}
Figure \ref{MDevo} shows the mass--density distributions of both the rocky (top row) and icy core (bottom row)  populations at 5 Gyr for different maximal distances from the star.  Planets that have lost all H/He lie along  the line labeled as ``Earth like'' in the top row, and somewhat above the ``Pure ice'' line in the bottom row. These extreme composition of bare planetary cores can (in this idealized case) be easily distinguished by their location in the mass--density space.

Below these lines, the evaporation valley is visible as a depletion of planets. Analogous to its different location in the $a$--$R$ diagram discussed in Section \ref{rockyicy}, it is located here at densities of about 3-7 g/cm$^{3}$ (depending on mass) for the rocky core population, but at 1-3 g/cm$^{3}$ in the icy core population. The red colors make it clear how strongly the envelope masses of close-in planets of a few $\mearth$ get reduced relative to their post-formation values. In contrast to these low-mass, gas giants only lose a few percent of their initial envelope by evaporation, visible form the blue colors of giant planets \citep[e.g.,][]{Tian2005,Murray-Clay2009,Owen2012,Jin2014}.

\begin{figure*}
 \includegraphics[width=17.5cm]{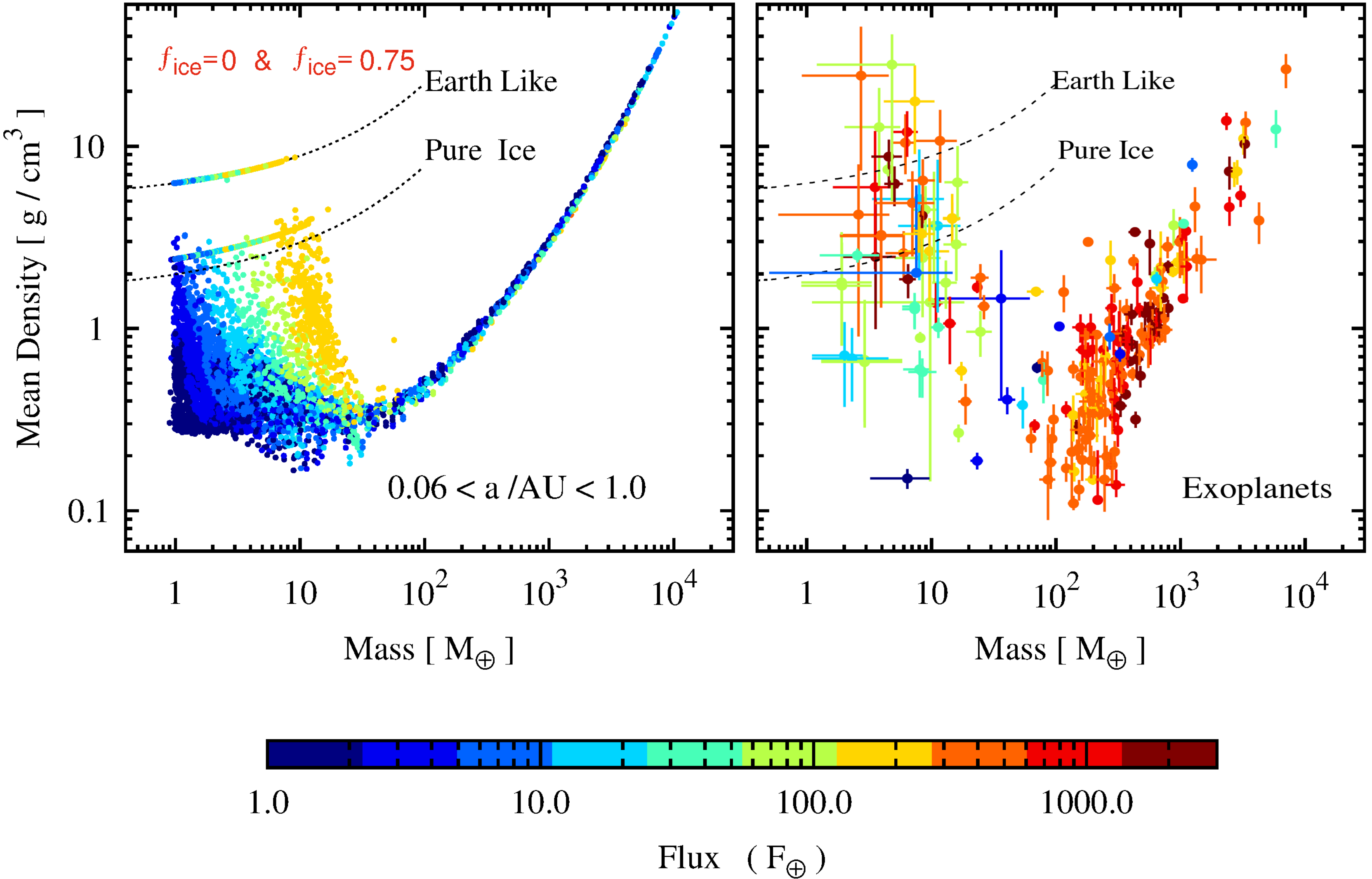}
 \caption{Comparison of the mass--density distributions of the synthetic planet populations and the known exoplanets. The left panel shows the combined synthetic rocky and icy core  populations. The right panel shows the exoplanets, compiled from www.exoplanets.org, \citet{Marcy2014} and \citet{Lissauer2013}. The color of each point shows the incident flux of a planet relative to the flux that the Earth receives from the Sun ($F_{\oplus}$). At low masses, more irradiated planets tend to have a higher density (a consequence of atmospheric escape), whereas for giant planets, more irradiated planets tend to have a lower density (a consequence of bloating).}
  \label{MDcomp}
\end{figure*}

The evaporation valley can be occupied if we add planets with orbital distances $>$ 2 AU (right column of Fig. \ref{MDevo}), where evaporation has a weak influence on planetary evolution so that low-mass planets can retain at least a portion of their (tenuous) H/He envelopes. Adding such distant planets also populates the bottom left corner where planets cannot retain their envelopes if only smaller orbital separations are considered. We thus see that low-mass planets with lower densities should get detected as observations allow to determine the densities of planets at increasingly large distances. 

From the mass--density plot we see that if the actual planetary population would consist of planets with rocky, icy, and mixed compositions, no clear valley would appear, as discussed already in the context of the radius distribution in Sect. \ref{rockyicy}. In view of the observations of \citet{Fulton2017}, this seems however not to be the case.

\subsection{Comparison with the observed mass--density diagram}
Figure \ref{MDcomp} compares the mass--density distributions of the planets inside of 1 AU in the rocky and the icy populations with the known exoplanets.

The color of each point in the figure indicates the incident flux on a planet, in comparison to the flux that the Earth receives from the Sun ($F_{\oplus}$). Note that the observed population includes many planets that are more strongly irradiated than the planets in the synthetic population where the maximum flux is about 1/$0.06^{2}\approx278 F_{\oplus}$. For the synthetic population, the colors again show that at given total mass, planets at higher fluxes have a higher density, and that the masses of the planets that can retain at least a part of its primordial H/He envelope increases with increasing incident flux. 

Most of the low-mass exoplanets in the right panel are from the Kepler satellite \citep{Marcy2014,Lissauer2013}. There are large error bars in the planetary masses, and hence the planetary mean densities. But the general trends found in the synthetic populations can still be found in the actual exoplanets. The mass--density distribution of the known exoplanets, first of all,  shows the same {characteristic} V-shape.  The mass with the lowest densities is at about 200 $\mearth$, but the minimum is very broad in mass, and affected by the bloating, making a more quantitative comparison with the synthetic population currently difficult.  

Furthermore, there is no planet $\lesssim10\mearth$ with a density less than 0.5 ${\rm g}~ {\rm cm}^{-3}$ inside of 0.1 AU (incident flux $>$ 100 $F_{\oplus}$), because planets at these distances are very sensitive to evaporation and can lose a large amount of their initial H/He envelopes during evolution. Such planets would contradict evaporation models. Many of the low-mass cores that receive a flux that $>$ 100 $F_{\oplus}$ lie between the mass--density curves of the Earth-like and pure-ice cores, suggesting that they may be bare cores without an envelope.
Roughly speaking, for giant planets, the mass--density distribution of the exoplanets is also similar to the gas giants in the synthetic  populations, although the distribution of densities of the actual giant exoplanets is not as sharp as the distribution in the synthetic populations, as expected.  

There is another interesting aspect shown in the mass--density distribution of the actual exoplanets. For giant planets, at a fixed planetary mass, those receiving high incident fluxes have a lower density {\citep[e.g.,][]{Laughlin2011}}. For low-mass planets, there is a hint that it is rather the opposite, i.e., those {that} receive high incident fluxes have a larger density. This shows that for giant planets, bloating is the dominating effect produced by the strong incoming flux {\citep[e.g.,][]{Thorngren2017}}. But for low-mass planets, the dominant effect of intense irradiation is atmospheric escape, which increases the planetary density such that density and orbital distance is anti-correlated for low-mass planets

\section{Summary and Conclusions}\label{discussion}
We have investigated how the population-wide statistical imprints of atmospheric escape depend on the bulk composition of the planetary cores using an end-to-end model of planet formation and evolution.  We have found that the location of the ``evaporation valley"  in the two-dimensional distance--radius plane and the associated one-dimensional bimodal distribution of radii \citep{Owen2013,Lopez2013,Jin2014,ChenRogers2016} clearly differ depending on the ice mass fraction of the planetary cores. Thus, we can use the imprints of evaporation  to break the compositional degeneracy existing otherwise in the mass--radius relationship of close-in low-mass exoplanets.  As the most important result we have found, by comparing model and observations, that the location of the gap both in the distance--radius plane and {in} the radius distribution recently found in the Kepler data \citep{Fulton2017} is consistent with a predominantly {Earth-like} rocky composition of the cores, but inconsistent with a mainly icy composition. In more details, we have  addressed this problem from three perspectives:

\subsection{{The locus of the valley}}
In the first part of the paper, in Sect. \ref{rockyicy},  we have studied the location of the evaporation valley and the associated minimum in the radius distribution with synthetic planet populations. 

Close-in, low-mass planets can be quickly evaporated to bare cores during the evolution phase after formation due to their low gravities and the strong incoming stellar XUV flux. These low-mass bare cores are well separated from the planets that retain at least a portion of their primordial H/He envelopes.

As a result, an ``evaporation valley" of $\sim$0.5 $R_{\oplus}$ in width underpopulated with planet forms, which runs diagonally downward in the $a$--$R$ distribution \citep{Owen2013,Lopez2013,Jin2014,LopezRice2016}. As a consequence, the one-dimensional radius distribution becomes bimodal, with the minimum  corresponding to the distance-weighted depletion of planets. These prominent evaporation features are not very sensitive to the loss efficiency in an evaporation model \citep{Jin2014}.

We have studied the location of this ``evaporation valley'' and the minimum in the radius distribution in two synthetic  populations that only differ by the composition of the solid core. In the first population all solid cores have an Earth-like rocky composition. In the second all cores contain 75\% of ice in mass, as expected for a formation outside of the water iceline. All planets start  with primordial H/He given by the formation model.

The ``evaporation valley" in the rocky core population occurs at about $R_{\rm bare,rocky}\approx1.6 \times (a/0.1$AU$)^{-0.27} R_{\oplus}$ in the $a$--$R$ plane (Figure \ref{twopop}) in agreement with the models of \citet{LopezRice2016}, and the associated minimum in the 1D radius distribution is centered around 1.6 $\rearth$ (Fig. \ref{compkeplernew}). In the icy core population, the valley is at about $R_{\rm bare,icy}\approx2.3 \times (a/0.1$AU$)^{-0.27} R_{\oplus}$, and the minimum is centered around 2.4 $\rearth$.  The reason for this difference is that a large amount of ice in a core decreases the core density (by about a factor 2, Fig. \ref{densfice}), which first makes the planets with icy cores more vulnerable to evaporation when they still have H/He, and second leads to larger radii of the bare cores once the H/He is evaporated. The different locations of the evaporation imprints mean that in a population with mixed core compositions, the evaporation features would be blurred or even removed\footnote{Before the publication of the \citet{Fulton2017} study we would have argued that the absence of clear evaporation imprints in the older Kepler data analyses \citep[e.g.,][]{Petigura2013} indicate mixed rocky and icy core compositions. }.

We have then compared the location of the valley in the $a$--$R$ plane and the minimum in the radius distribution of the two synthetic populations with the observational counterparts recently found in the Kepler data  \citep{Fulton2017}. As the most important result of this study, we have found that the  imprints of evaporation in the rocky core population are consistent with observations, but not in the ice cory population (Fig. \ref{twopop}). In the rocky core population, the ``evaporation valley'' in the $a$--$R$ plane occurs at a similar location as in the observations, whereas in the icy core population it occurs at radii that are about 0.7-1 $\rearth$ too large. Also the associated location of the minimum in the {1D} radius histogram in the rocky core population agrees with the observed location at about 1.7 $\rearth$ \citep[][]{Fulton2017}. The minimum in the icy core population is in contrast again at too large radii. In the icy core population the minimum even occurs quite exactly at the position of the observed sub-Neptune maximum of \citet{Fulton2017}. This makes this population clearly inconsistent with observations (Fig. \ref{compkeplernew}).

If the observed gap is really due to evaporation, we can conclude from this comparison that the cores of close-in low-mass Kepler planets are predominantly made of silicates and iron, without large amounts of ices. From a formation point of view it seems rather unlikely that other effects like a late gas-poor formation \citep{Lee2014,LopezRice2016} or envelope removal by giant impacts \citep[e.g.,][]{Schlichting2015} should not have played a role as well. Our study shows the consequences of evaporation only, allowing to infer the differences to observations and the possible effects of other envelope removal mechanisms. 
Our statistical results also does not exclude that some close-in, low-mass planets still have a large ice content. But this should not be the dominant composition. Recently, \citet{Lopez2016} also reached the conclusion of rocky core compositions from an analysis of a different aspect, the radii of ultra-short-period planets. The location of the valley in the rocky population is also compatible with the transition at $\approx$1.6 $\rearth$  found by \citet{Rogers2015}. 

{It is interesting to note that spectroscopic observations of polluted white dwarfs indicate a dry Earth-like bulk elemental composition for most accreted asteroids and minor planets \citep{Jura2014,Xu2014}. The observed oxygen abundances show that the polluted WD viewed as an ensemble accreted dry material where water is at most a few percent of the total accreted mass (but exceptions exist). Even if the WD and the Kepler planets studied here probe different evolutionary stages of planetary systems, these findings point to the same consistent direction of roughly Earth-like bulk compositions without much water.} 

The rocky composition suggests that these planets have accreted mainly inside of the water iceline. Combined with the population-wide imprints of  orbital migration in the Kepler data like the frequency maxima outside of MMR period ratios \citep{Fabrycky2014}, the picture arises that orbital migration in the protoplanetary disk was important for the formation of these planets, but that  migration was confined to the inner disk. 

The region in the $a$--$R$ plane containing bare planets that have lost all primordial H/He forms in a log-log plot (Fig. \ref{twopop}) a triangle. Therefore, we call this region the ``triangle of evaporation''. It is an interesting region, because the degeneracy of possible planetary compositions for a given mass and radius is reduced here. 

\subsection{{Compositions in the triangle of evaporation}}
In the second part of the paper (Sect. \ref{sect:icemassfraction}) we have tried to statistically infer the fraction of planets in the triangle of evaporation containing a high ice-mass fraction among the planets there with a known density. From the first part of the paper, we expect that most planets in this region should have a rocky composition. Finding that most planets in the triangle of evaporation with known density would require large amounts of ice would  be a contradiction. 

For this, we have analyzed 55 planets from the \citet{WeissMarcy2014} sample. Given the mass and radius and their 1-$\sigma$ errors, and under the assumption that the planets have below a possible ice layer an (approximately) Earth-like silicate:iron ratio, we have calculated with interior structure models \citep{Mordasini2012b} the amount of ice that is necessary to explain their density. We have conducted four statistical analyses, combining two different equations of state (\citealt{Seager2007} and \citealt{Grasset2009}), with two assumptions about the impact of a low-density vapor layer. In one, the NoDampf analysis, the effect of such a low-density layer {was} neglected. In the other, the Dampf analysis, we have subtracted the thickness of the layer from the radius, while neglecting its mass. One finds that the general statistical trends are comparable in all four analyses.  

The derived ice mass fraction combined with the position of a planet either inside or outside of the triangle of evaporation for rocky cores (Eq. \ref{eq:Rbarrocky}) allows to classify the 55 planets in 8 types (Table \ref{tablefice}). The  number of planets identified are:  Type 1: outside, with H/He: 11 planets. Type 2: outside, rocky: 0. Type 3: outside, with volatiles (unconstrained whether ice and/or H/He): 14. Type 4: outside, unconstrained:  6. Type 5: inside, with H/He: 0. Type 6: inside, rocky: 7. Type 7: inside, icy: 3. Type 8: inside, unconstrained: 15. A closer look at the three Type 7 planets identified in the NoDampf analysis shows that none of them has a very secure water-dominated composition \citep[cf.][]{Dorn2017b,Lopez2016}. In the Dampf analysis, the composition of all these three planets is even unconstrained. 

The absence of Type 2 planets means that no rocky planets outside of the triangle of evaporation were identified.  Type  5 planets were neither found. {These} would be planets inside of the triangle of evaporation that need H/He to explain their low density. The absence of these two types is in agreement with a scenario where planets in the triangle of evaporation lose the H/He, while those outside start with H/He and keep it.

It is interesting to {study} the derived compositional types in the $a$--$R$ plane (Figs. \ref{RcompoNoAtmoSubtr} and \ref{RcompoAtmoSubtr}). A clear compositional gradient with increasing planet radius is seen, similar to earlier studies \citep[e.g.,][]{Marcy2014,Rogers2015,WolfgangLopez2015}: for $R\lesssim 1.6 \rearth$, we find rocky compositions. For 1.6 to 3 $\rearth$, volatiles are required, but it is  unconstrained whether it is H/He and/or ices. Finally, for $R\gtrsim 3 \rearth$, H/He is usually required to explain a planet's density. The theoretically predicted transition from rocky planets to those with H/He given by the evaporation valley agrees in a general way  with the observed  transition, but the small sample size and the large observational error bars make it difficult to make more precise statements.

The most important question we wanted to address in the second part was whether there are many clearly ice-dominated planets in the triangle of evaporation. Tables \ref{statsnodampf} and \ref{statsdampf}  summarize the number and percentages of the different planet types. They shows that taken at face value, for the planets with a constrained composition, between 70-100\% of the planets in the triangle of evaporation have a rocky composition, and 0-30\% an icy composition. The actual value is probably closer to zero. It is clear that these values are derived from a small sample, with a fixed iron:silicate ratio, and a simple analysis of the errors using just the 1-$\sigma$ uncertainties instead of a full Bayesian analysis \citep[e.g.,][]{Rogers2015,Dorn2017a}. But they nevertheless hint at a  predominantly rocky composition of planets in the triangle of evaporation. Based on the densities, we have thus found an agreement with the statistical result on the location of the evaporation valley from the first part that is based on radii only.

{\subsection{The mass--mean density diagram in time}}
In the last part (Section \ref{massdensity}) we have studied the planetary mass--mean density diagram as a function of distance and time. We find that the mass--density distribution of a planet population contains  important information both about planet formation and evolution  \citep{Rauer2013,HatzesRauer2015,Baruteau2016}.

{T}he general structure of the synthetic $M$--$\bar{\rho}$ diagram (Figure \ref{MDoverview}) {is a characteristic broken} V-shape. The left branch of the ``V'' {consists of solid planets and, separated from then by the evaporation valley,} low-mass core-dominated planets with H/He. For them, the most distinct feature is that their location in the mass-density plot reveals their envelope mass fraction \citep{Lopez2014}. This part of the V-shape also shows the effect of evaporation for close-in low-mass planets: at a given total mass, the hotter a planet, the higher its density since more primordial H/He was lost due to stronger evaporation. This indicates that mainly evolution in the form of evaporation shapes the radii at least for small orbital distances \citep{Owen2013}. 

Another notable feature in the $M$--$\bar{\rho}$ diagram is a change of regime at about $\sim$ 30 $M_{\oplus}$. At this mass, for planets with (remaining) primordial H/He, the density changes from decreasing with increasing mass because of an increasing H/He mass fraction, to a density that is first only weakly dependent on mass (for $M\lesssim 70 M_\oplus$), to finally a density that  increases with mass, because of the increasing self-compression. The lowest mean density are planets of 10-30 $M_{\oplus}$. The right part of the V-shape consists of gas-dominated giant planets.

We have studied the evolution of the mean density in time (Fig. \ref{MDtemp}). This is particularly important when considering that the  PLATO 2.0 mission can  determine the ages of the host stars and observe the temporal evolution of planets. As expected, the mean densities of  planets with H/He increases in time. For the giant planets this is mostly due to cooling and contraction at constant mass. It causes the densities of gas-dominated planets to increase by about a factor 2-3 from 20 Myr to 10 Gyr.  For close-in core-dominated planets with H/He, evaporation is in contrast the dominant effect shaping the density in the interval of orbital distance that we have studied. Evaporation removes close-in low-mass  planets with low density in the mass--density space, mainly in the first 100 Myr after formation.  This can lead to an increase of the mean density by up to a factor $\sim$100.

No significant change of the mean density in time occurs for planets without H/He. Determining observationally whether the mean density of a certain sub-group of planets (for example in a interval in mass and insolation) changes between ages of 0.1 and 5 Gyr thus allows to constrain whether they contain H/He. 

A comparison of the synthetic and the observed mass--density diagram (Fig. \ref{MDtemp})  shows that the distribution of the known exoplanets also has a similar V-shape as the synthetic population. It also seems to be consistent with a similar turning point, but it is difficult to pinpoint it exactly because of the bloated giant planets. 

Furthermore, for observed giant planets, at a fixed mass, those receiving high incident fluxes tend to have a lower density {\citep[e.g.,][]{Laughlin2011}}.  For observed low-mass planets, there is a hint that it is rather the opposite, i.e., that those receiving a higher fluxes have a higher density. In the synthetic population, this correlation is very clear. This shows that for giant planets, bloating is the dominating effect caused by the strong incoming flux {\citep[e.g.,][]{Thorngren2017}}. But for low-mass planets, the dominant effect of intense irradiation is atmospheric escape,   such that density and orbital distance is anti-correlated. 

\subsection{{Outlook}}
Coming back to the valley of evaporation, TESS \citep{Ricker2010}, CHEOPS \citep{Broeg2013}, and PLATO 2.0 {\citep{Rauer2013}} will yield accurate radii and RV follow-up or TTVs masses of planets in the $a$-$R$ parameter space on both sides of the  valley. This will allow to much better understand the various compositional transitions that are currently very difficult to pinpoint for individual planets because of the large error bars. An important, currently open question is how the transition from solid planets to planets with H/He depends on orbital distance. This should allow to disentangle different mechanisms like evaporation or impacts \citep{LopezRice2016}. It will additionally be interesting to see whether spectroscopic observations {and observations of escape} find an associated transition in the atmospheric properties, for example in terms of the mean molecular weight {or the escape rate}. This will allow to understand how bulk and atmospheric composition correlate.

Planetary evolution can sometimes blur the imprints of the formation epoch. Here we could instead have a positive opposite situation: For close-in low-mass planets, the mass distribution is under evaporation continuous without gap or local minimum, in contrast to the radius distribution. The reason is that the primordial H/He mass fraction of these planets is so small compared to the total mass that its loss does not  significantly reduce the total mass \citep{Jin2014}. This means that the mass distribution reflects formation, whereas the radius distribution shaped by evaporation mainly reflects evolution. But interestingly, the evolutive imprint of evaporation allows to better understand formation by revealing indirectly the core composition via the location of the valley of evaporation.

{\textit{Notes added.} After the submission of this paper we became aware of the work of \citet{Owen2017} who had independently reached the same main conclusions as we do in the present work regarding the composition of the Kepler planets. Comparison of their Figure 9 with our Figure 2 shows that the two papers agree well regarding the predicted location of the valley, with differences of about 0.2  $\rearth$ or less.}

{In a recent observational study, \citet{vaneylen2017} report a negative  slope of the occurrence valley with orbital distance, consistent with the predictions of atmospheric escape. }

\section*{Acknowledgments}
S. J. acknowledges the support of the National Natural Science Foundation of China (Grants No. 11273068, 11473073, 11503092, 11661161013, 11773081), the innovative and interdisciplinary program by CAS (Grant No. KJZD-EW-Z001) and the Foundation of Minor Planets of the Purple Mountain Observatory. C.M. acknowledges the support from the Swiss National Science Foundation under grant BSSGI0$\_$155816 ``PlanetsInTime''. Parts of this work have been carried out within the frame of the National Center for Competence in Research PlanetS supported by the SNSF. {We thank an anonymous referee for a helpful report.}

%\bibliographystyle{apj} % style aa.bst 
%\bibliography{paper2.bbl}

\end{document}